\documentclass[11pt,a4paper]{article}
\pdfoutput=1 

\usepackage{jheppub}

\usepackage[T1]{fontenc}

\usepackage[latin9]{inputenc}
\usepackage{esint}

\newcommand{\noun}[1]{\textsc{#1}}
\newcommand{\POWHEG}{\noun{Powheg}}
\newcommand{\POWHEGBOX}{\noun{Powheg Box}}
\newcommand{\MINLO}{\noun{Minlo}}
\newcommand{\HJMINLO}{\noun{Hj-Minlo}}
\newcommand{\HNNLO}{\noun{Hnnlo}}
\newcommand{\NNLOPS}{\noun{Nnlops}}
\newcommand{\NLOPS}{\noun{Nlops}}
\newcommand{\PYTHIA}{\noun{Pythia}}
\newcommand{\MCatNLO}{\noun{MC@NLO}}
\newcommand{\ggH}{\noun{H}}
\newcommand{\HQT}{\noun{HqT}}
\newcommand{\HJ}{\noun{Hj}}
\newcommand{\MEPS}{\noun{Meps}}
\newcommand{\JETVHETO}{\noun{JetVHeto}}
\newcommand{\FASTJET}{\noun{FastJet}}
\newcommand{\as}{\alpha_{\scriptscriptstyle \mathrm{S}}}
\newcommand{\Kr}{K_{\scriptscriptstyle \mathrm{R}}}
\newcommand{\Kf}{K_{\scriptscriptstyle \mathrm{F}}}
\newcommand{\mur}{\mu_{\scriptscriptstyle \mathrm{R}}}
\newcommand{\muf}{\mu_{\scriptscriptstyle \mathrm{F}}}
\newcommand{\pt}{p_{\scriptscriptstyle \mathrm{T}}}
\newcommand{\pth}{p_{\scriptscriptstyle \mathrm{T}}^{\scriptscriptstyle \mathrm{H}}}
\newcommand{\ptjone}{p_{\scriptscriptstyle \mathrm{T}}^{\scriptscriptstyle \mathrm{j_{1}}}}
\newcommand{\kt}{k_{\scriptscriptstyle \mathrm{T}}}
\newcommand{\mh}{m_{\scriptscriptstyle \mathrm{H}}}

\newcommand{\hc}{\beta}
\newcommand{\hgam}{\gamma}

\usepackage[mathscr]{euscript}

\preprint{\\\\MCnet-13-11\\CERN-PH-TH/2013-205\\OUTP-13-18P}

\title{{NNLOPS simulation of Higgs boson production}}

\author[a,b]{Keith Hamilton}
\author[c]{Paolo Nason}
\author[d]{Emanuele Re}
\author[d]{Giulia Zanderighi}

\affiliation[a,1]{Department of Physics and Astronomy, University College London,\\London, WC1E 6BT, UK\note{From 1st August 2013.}}
\affiliation[b]{Theory Division, CERN,\\CH--1211, Geneva 23, Switzerland}
\affiliation[c]{INFN, Sezione di Milano Bicocca,\\Piazza della Scienza 3, 20126 Milan, Italy}
\affiliation[d]{Rudolf Peierls Centre for Theoretical Physics, University of Oxford\\1 Keble Road, UK}

\emailAdd{keith.hamilton@ucl.ac.uk}
\emailAdd{paolo.nason@mib.infn.it}
\emailAdd{e.re1@physics.ox.ac.uk}
\emailAdd{g.zanderighi1@physics.ox.ac.uk}

\abstract{We detail a simulation of Higgs boson production via gluon
  fusion, accurate at next-to-next-to-leading order in the strong
  coupling, including matching to a parton shower, yielding a
  fully exclusive, hadron-level description of the final-state. The
  approach relies on the \POWHEG{} method for merging the NLO Higgs
  plus jet cross-section with the parton shower, and on the \MINLO{}
  method to simultaneously achieve NLO accuracy for inclusive Higgs
  boson production. The NNLO accuracy is reached by a reweighting
  procedure making use of the \HNNLO{} program.}


\begin{document}
\maketitle
\flushbottom

\section{Introduction}

Presently, the discovery of a new spin-zero particle in the search
for the Standard Model (SM) Higgs boson at the Large Hadron Collider
(LHC), 
by the ATLAS and CMS collaborations \cite{Aad:2012tfa,Chatrchyan:2012ufa},
has focused the physics agenda on studying its properties, to search
for possible departures from the SM predictions. The accurate measurement
of the couplings and quantum numbers of the new particle forms the
basis of this endeavour. Thus far, with present data, the newly discovered
boson shows no significant deviations from the SM expectations.

One of the fundamental limitations in our ability to study the nature
of the new particle is the precision afforded by our theoretical predictions
in describing its production and decay rates, and the associated experimental
acceptances. Now, following the initial discovery, with the mass of
the would-be Higgs boson known, it is reasonable to expect that the
associated data will grow at a considerable rate when the LHC restarts
in 2014/15. It is foreseeable then, that as the statistical errors
on the experimental measurements diminish, the accuracy of
the theoretical predictions will become an even more pressing issue,
as work continues towards constraining deviations from the apparent
SM behaviour.

The anticipation of new and exciting discoveries at the LHC has catalysed
significant progress in addressing concerns regarding the precision
of theoretical predictions (for a recent review see e.g. Ref.~\cite{Salam:2012cv}
and references therein). Notably, the last ten years have witnessed
remarkable developments in precision fixed order QCD calculations
for the main (gluon fusion) Higgs boson production channel, beginning
with next-to-next-to-leading order (NNLO) predictions for the total
inclusive cross section \cite{Harlander:2002wh,Anastasiou:2002yz,Ravindran:2003um},
followed by analogous computations of differential quantities \cite{Anastasiou:2005qj,Catani:2007vq}.
Most recently, landmark calculations have been carried out in deriving
the full or partial NNLO QCD corrections to a number of $2\rightarrow2$
partonic scattering processes \cite{Czakon:2012zr,Czakon:2012pz,Ridder:2013mf,Czakon:2013goa},
including $gg\rightarrow H+\mathrm{jet}$ \cite{Boughezal:2013uia}.

While the emergence of fully differential fixed order NNLO calculations
represents an important theoretical breakthrough, the practical value
of which is underlined by their use in the Higgs boson discovery analyses
\cite{Aad:2012tfa,Chatrchyan:2012ufa}, their description of the final
state is comprised of at most two QCD partons. Multiple parton emission,
resummation effects, hadronisation and the underlying event are not
accounted for. All of the latter class of contributions represent
corrections formally beyond NNLO for inclusive quantities. They are
however very relevant for the exclusive description of the final state.
Furthermore, depending on the nature of the observables under consideration,
they may sum to give corrections as significant as the NNLO ones,
or supersede them altogether, in kinematic regions where fixed order
perturbation theory becomes unreliable.

In parallel with advances in fixed order computations, significant
progress has been made on the description of exclusive final states, most notably
through the development of methods for the consistent inclusion of
next-to-leading order (NLO) matrix elements in parton shower Monte
Carlo event generators (\NLOPS{}) \cite{Frixione:2002ik,Nason:2004rx}.
In less than ten years since their inception, these matching schemes
have evolved to the point where
the construction of \NLOPS{} simulations, for even complex
multi-leg processes, proceeds with a high level of automation
\cite{Alioli:2010xd,Frederix:2011zi,Platzer:2011bc,Hoeche:2012ft},
with little or nothing required in the form of \NLOPS{} expertise.
Like the fixed order NNLO computations, the value of \NLOPS{} simulations
is well established and they too have been used in the Higgs boson
discovery analyses.

The rapid development in \NLOPS{} techniques in the last three years
has led to the creation of a number of public codes for most processes
of interest at the LHC, among which those of the form $pp\rightarrow X+n\,\mathrm{jets}$,
with $X$ being e.g. a $W/Z/H$ boson \cite{Alioli:2010qp,Frederix:2011ig,Hoeche:2012ft,Re:2012zi,Campbell:2012am,Campbell:2013vha}.
In view of the positive experience with matrix-elements and parton
shower merging schemes (\MEPS{}) \cite{Mangano:2001xp,Catani:2001cc,Lonnblad:2001iq,Alwall:2007fs},
several efforts have appeared in the literature towards the developments
of methods that allow the merging of NLO samples with different associated
jet multiplicities~\cite{Lavesson:2008ah,Alioli:2011nr,Hamilton:2012np,Gehrmann:2012yg,Hoeche:2012yf,Frederix:2012ps,Lonnblad:2012ng,Platzer:2012bs,Lonnblad:2012ix,Alioli:2012fc,Hamilton:2012rf,Hartgring:2013jma}.
We now focus upon the example of inclusive Higgs and Higgs plus one
jet generator (henceforth \ggH{} and \HJ{}). Ideally, the merging
procedure should yield a generator that is NLO accurate for both fully
inclusive observables, and for observables requiring the presence
of one associated jet. Recall that the \ggH{} generator is NLO accurate
for inclusive observables but only LO accurate in the description
of the associated jet, while the \HJ{} generator is NLO accurate
in the description of the associated jet, but it cannot be used to
compute fully inclusive observables. It is clear that the construction
of a `merged' generator of this kind is a first step in the 
development of a NNLO accurate generator matched to a parton shower.
In fact, the latter is better than the former only for inclusive quantities,
where the former is only NLO accurate.

Among the new multi-jet merging schemes the \MINLO{} approach, as
it is set out in \cite{Hamilton:2012rf}, is unique in that it does
not resort to the introduction of an unphysical \emph{merging scale}
cut to partition the phase space into a 0-jet region, to be populated
by the H \NLOPS{} simulation, and a $\ge1$-jet region, to be filled
by the \HJ{} one. On the contrary, the \MINLO{} recipe acts to
extend the reach of the underlying \HJ{} NLO computation so as to
return NLO accurate predictions for observables inclusive with respect
to all QCD radiation, without the introduction of any additional unphysical
parameters. As a consequence, the \HJMINLO{} event generator provides
all of the same fixed order accuracy as a fully differential NNLO
computation, except fully inclusive observables, for which it is only
NLO accurate. In ref.~\cite{Hamilton:2012rf} it was argued that,
by a simple reweighting procedure, full NNLO accuracy can actually be
achieved with such a generator. In the present work we implement
this reweighting,
yielding a first true \NNLOPS{} simulation for Higgs boson production.
The fact that the standalone \HJMINLO{} simulation already achieves
NLO accuracy for inclusive Higgs boson production observables is at
the heart of this development. Were it not for this feature, the magnitude
of the weights required to correct the inclusive distributions to
NNLO accuracy would be such that the NLO precision for \HJ{} observables
is then lost.

The remainder of this paper is structured as follows. In section~\ref{sec:Theoretical-framework}
we present the theoretical framework underlying the \NNLOPS{} method,
starting with the basic formulation in section~\ref{sub:NNLOPS},
followed by simple refinements in section~\ref{sub:Simple-variations}.
Section~\ref{sec:Estimating-uncertainties} describes our prescription
for the determination of theoretical uncertainties in the new method.
In section~\ref{sec:Results} we present a selection of results obtained
having implemented our \NNLOPS{} method, with the aim of probing
and validating the procedure. Lastly, in section~\ref{sec:Conclusions},
we present our conclusions and comment on further developments.

\section{Theoretical framework \label{sec:Theoretical-framework}}

In ref.~\cite{Hamilton:2012rf} it was proven that the \HJMINLO{}
computation is NLO accurate for Higgs plus one jet (\HJ{}) and Higgs
(\ggH{}) inclusive observables. As noted in the introduction and
outlined in ref.~\cite{Hamilton:2012rf}, this property is crucial
for the promotion of the \HJMINLO{} simulation from \NLOPS{} to
\NNLOPS{}. In this section we describe in detail the procedure
for reaching NNLO accuracy. For details regarding how the \HJMINLO{}
simulation first attains NLO accuracy for inclusive Higgs production
observables we refer the reader back to ref.~\cite{Hamilton:2012rf}.

\subsection{Method\label{sub:NNLOPS}}

Let us abbreviate by $d\sigma^{{\scriptscriptstyle \mathrm{MINLO}}}$
the cross section obtained from the \HJMINLO{} event generator,
fully differential in the final state phase space, $\Phi$, at the
level of the hardest emission events, i.e. prior to showering. On
integration, this distribution reproduces the leading order, $\mathcal{O}\left(\as^{2}\right)$,
and next-to-leading order contributions, $\mathcal{O}\left(\as^{3}\right)$,
in the perturbative expansion of the Higgs boson rapidity distribution.
In addition, unenhanced spurious terms entering at $\mathcal{O}\left(\as^{4}\right)$
and higher are present, since the fully differential cross section
includes all order contributions in the Sudakov form factors, and
contributions of order $\mathcal{O}\left(\as^{4}\right)$ to the
Higgs plus jet cross section. By analogy
we denote the conventional fixed order, next-to-next-to-leading order
differential distribution by $d\sigma^{{\scriptscriptstyle \mathrm{NNLO}}}$.
Since the rapidity distributions derived from $d\sigma^{{\scriptscriptstyle \mathrm{MINLO}}}$
and $d\sigma^{{\scriptscriptstyle \mathrm{NNLO}}}$ are formally identical
up to and including $\mathcal{O}\left(\alpha_{{\scriptscriptstyle \mathrm{S}}}^{3}\right)$
terms, it follows that their ratio is equal to one up to $\mathcal{O}\left(\alpha_{{\scriptscriptstyle \mathrm{S}}}^{2}\right)$
terms: 
\begin{eqnarray}
\mathcal{W}\left(y\right) & = & \frac{\smallint d\sigma^{{\scriptscriptstyle \mathrm{NNLO\phantom{i}}}}\,\delta\left(y-y\left(\Phi\right)\right)}{\smallint d\sigma^{{\scriptscriptstyle \mathrm{MINLO}}}\,\delta\left(y-y\left(\Phi\right)\right)}\label{eq:NNLOPS-proof-W-is-1-plus-aS2-i}\\
 & = & \frac{c_{2}\alpha_{{\scriptscriptstyle \mathrm{S}}}^{2}+c_{3}\alpha_{{\scriptscriptstyle \mathrm{S}}}^{3}+c_{4}\alpha_{{\scriptscriptstyle \mathrm{S}}}^{4}\phantom{+\ldots}}{c_{2}\alpha_{{\scriptscriptstyle \mathrm{S}}}^{2}+c_{3}\alpha_{{\scriptscriptstyle \mathrm{S}}}^{3}+c_{4}^{\prime}\alpha_{{\scriptscriptstyle \mathrm{S}}}^{4}+\ldots}\label{eq:NNLOPS-proof-W-is-1-plus-aS2-ii}\\
 & = & 1+\frac{c_{4}-c_{4}^{\prime}}{c_{2}}\,\alpha_{{\scriptscriptstyle \mathrm{S}}}^{2}+\ldots,\label{eq:NNLOPS-proof-W-is-1-plus-aS2-iii}
\end{eqnarray}
where the $c_{i}$ are simply constant $\mathcal{O}\left(1\right)$
coefficients.

While it is obvious that by this reweighting the inclusive rapidity
distribution acquires NNLO accuracy, the crucial point here is that
the NLO (i.e. ${\cal O}(\as^{4})$) accuracy of the cross section
in the presence of jets (that starts at order $\as^{3}$) is maintained,
since the reweighting factor combined with this cross section
yields spurious terms of order $\as^{5}$
and higher.
We stress again that, were it not for the special property that the
\noun{Hj-Minlo} generator reproduces the conventional fixed order
result up to and including NLO terms, $\mathcal{W}\left(y\right)$
would yield relative corrections of $\mathcal{O}(\as)$, thus spoiling the
NLO accuracy of Higgs plus one jet distributions.

We shall now demonstrate that the \HJMINLO{} generator reweighted
with the procedure outlined above achieves $\mathcal{O}(\as^{4})$
accuracy for all observables. To begin with we must prove the
following theorem: 
\begin{quote}
\textit{A parton level Higgs boson production generator that is
accurate at ${\cal O}(\alpha_{{\scriptscriptstyle \mathrm{S}}}^{4})$ 
for all IR safe observables that vanish with the transverse momenta
of all light partons, and that also reaches 
${\cal O}(\alpha_{{\scriptscriptstyle \mathrm{S}}}^{4})$
accuracy for the inclusive Higgs rapidity distribution,
achieves the same level of precision for all IR safe observables,
i.e. it is fully NNLO accurate.} 
\end{quote}
To this end, we consider a generic observable $F$ that is an infrared
safe function of the final state kinematics. Its value will be given
by 
\begin{equation}
\langle F\rangle=\int d\Phi\,\frac{d\sigma}{d\Phi}\, F(\Phi),\label{eq:NNLOPS-F-i}
\end{equation}
with a sum over final state multiplicities being implicit in the
phase space integral. Infrared safety ensures that $F$ has a
smooth limit when the transverse momenta of the light partons
vanish. Such a limit may only depend upon the Higgs boson's
rapidity, $y$, since it is the only observable left when no other
partons are resolved. We generically denote such a limit by $F_{y}$.
The value of $\langle F\rangle$ can be considered as the sum of two
terms: $\left\langle F-F_{y}\right\rangle + \left\langle F_{y}\right\rangle $.
Since, $F-F_{y}$ tends to zero with the transverse momenta of
all the light partons, by hypothesis its value is
given with ${\cal O}(\alpha_{{\scriptscriptstyle \mathrm{S}}}^{4})$
accuracy by the parton level generator.
On the other hand,
\begin{equation}
\left\langle F_{y}\right\rangle =\int dy^{\prime}\,\frac{d\sigma}{dy^{\prime}}\,
F_{y}\left(y^{\prime}\right)\label{eq:NNLOPS-F-ii}\,,
\end{equation}
which is also exact at the ${\cal O}(\alpha_{{\scriptscriptstyle \mathrm{S}}}^{4})$
level by hypothesis. Thus, $\langle F\rangle=\left\langle F-F_{y}\right\rangle
+\left\langle F_{y}\right\rangle $ is accurate at the
${\cal O}(\alpha_{{\scriptscriptstyle \mathrm{S}}}^{4})$ level, proving
our theorem.

This theorem is easily generalized to arbitrary processes. It is enough
to replace $F_{y}$ in the above proof with $F_{\Phi_{B}}$, where $\Phi_{B}$
are infrared safe quantities parametrising the associated Born phase space.

The \HJMINLO{} parton level generator is, by itself, one that fulfills
the first condition required by our theorem; predicting IR
safe observables that vanish when the transverse momentum of the light
partons vanishes with $\mathcal{O}(\as^{4})$ accuracy.
In fact, this would be the case even without the \MINLO{} improvement.
The second hypothesised statement
entering the theorem, regarding NNLO accuracy of the Higgs boson's
rapidity spectrum, is realised by augmenting the  \HJMINLO{} 
generator by the reweighting procedure described above.
The proof of $\mathcal{O}(\as^{4})$
accuracy for these observables thus corresponds to the general proof
of NLO accuracy of the \POWHEG{} procedure, given in
refs.~\cite{Frixione:2007vw,Nason:2012pr}.

Observe, also, that for observables of the type
$\left\langle F-F_{y}\right\rangle$, adding the full shower development
does not alter the $\mathcal{O}(\as^{4})$ accuracy of the algorithm,
for the same reasons as in the case of the regular \POWHEG{} method.
The only remaining worry one can then have concerns the
possibility that
the inclusive Higgs boson rapidity distribution is modified by the parton
shower evolution at the level of $\mathcal{O}(\as^{4})$ terms. However, our
algorithm already controls the two hardest emissions with the
required $\as^{4}$ accuracy. A further emission from the shower is
thus bound to lead to corrections of higher order in $\as$.%
\footnote{Recall that \POWHEG{} simulations limit the transverse
momenta of branchings in the subsequent parton shower simulation to
be less than the hardest emission generated with respect to the 
underlying Born (passed through the so called \texttt{scalup} variable),
which here is the second emitted parton.%
}

We notice that the \NNLOPS{} event generator described here is NNLO
accurate in the same sense in which the current \MCatNLO{} or \POWHEG{}
type generators are NLO, i.e.~infrared safe observables
are NLO or NNLO accurate. The degree of logarithmic accuracy,
leading, next-to-leading or next-to-next-to-leading, is contingent
upon the way in which the associated Sudakov form factors are implemented.

It should be clear from our discussion that reweighting can be performed
at the partonic level either before or after the shower. It cannot
be carried out at the level of the generation of the underlying Born
configuration of the \HJMINLO{} generator, since the Higgs rapidity
changes after radiation.%
\footnote{The \POWHEGBOX{} implementation guarantees that the rapidity of
the system comprising the Higgs plus the hardest parton remains the
same after radiation, but not the rapidity of the Higgs itself.%
}

\subsection{Variant schemes\label{sub:Simple-variations}}

One can readily construct simple variants of the method discussed
in sect.~\ref{sub:NNLOPS}. In particular, rather than performing
a global reweighting of all \noun{Hj-Minlo} events using the fully
inclusive Higgs boson rapidity distribution, one can instead consider
splitting the cross section according to 
\begin{eqnarray}
d\sigma_{\phantom{0}} & = & d\sigma_{A}+d\sigma_{B}\,,\label{eq:NNLOPS-dsig-eq-dsig0-plus-dsig1}\\
d\sigma_{A} & = & d\sigma\, h\left(\pt\right)\,,\label{eq:NNLOPS-dsig0}\\
d\sigma_{B} & = & d\sigma\,\left(1-h\left(\pt\right)\right)\,,\label{eq:NNLOPS-dsig1}
\end{eqnarray}
where $\pt$ here represents some overall measure of the hardness
of radiation in the event, with $h$ a monotonic profile function
such that
$\lim_{p_{{\scriptscriptstyle \mathrm{T}}}\rightarrow0}\, h(\pt)=1$,
$\lim_{p_{{\scriptscriptstyle \mathrm{T}}}\gg m_{{\scriptscriptstyle \mathrm{H}}}}\, h(\pt)=0$,
and simply reweight the $d\sigma_{A}$ component rather than the
full cross section. A suitable
form for the profile function is 
\begin{equation}
h(\pt)=\frac{(\hc\, m_{{\scriptscriptstyle \mathrm{H}}})^{\hgam}}{(\hc\, m_{{\scriptscriptstyle \mathrm{H}}})^{\hgam}+\pt^{\hgam}},\label{eq:NNLOPS-hpT-fn-defn}
\end{equation}
where $\hc$ and $\hgam$ are constant parameters. We reweight the
\HJMINLO{} events with the factor 
\begin{equation}
\mathcal{W}\left(y,\, p_{{\scriptscriptstyle \mathrm{T}}}\right)=h\left(\pt\right)\,\frac{\smallint d\sigma_{A}^{{\scriptscriptstyle \mathrm{NNLO\phantom{i}}}}\,\delta\left(y-y\left(\Phi\right)\right)}{\smallint d\sigma_{A}^{{\scriptscriptstyle \mathrm{MINLO}}}\,\delta\left(y-y\left(\Phi\right)\right)}+\left(1-h\left(\pt\right)\right)\,.\label{eq:NNLOPS-overall-rwgt-factor}
\end{equation}
Multiplying the above equation by $d\sigma^{\mathrm{MINLO}}\delta(y-y(\Phi))$,
using equations~(\ref{eq:NNLOPS-dsig0}) and (\ref{eq:NNLOPS-dsig1}),
and integrating over the full phase space we obtain identically 
\begin{eqnarray}
\left(\frac{d\sigma}{dy}\right)^{{\scriptscriptstyle \mathrm{NNLOPS}}} & = & \left(\frac{d\sigma_{A}}{dy}\right)^{{\scriptscriptstyle \mathrm{NNLO}}}+\left(\frac{d\sigma_{B}}{dy}\right)^{{\scriptscriptstyle \mathrm{MINLO}}}\,.\label{eq:NNLOPS-NNLOPS-eq-NNLO_0+MINLO_1}
\end{eqnarray}
Eq.~(\ref{eq:NNLOPS-NNLOPS-eq-NNLO_0+MINLO_1}) differs from the
fixed order NNLO cross-section only by terms of order $\as^{5}$.

In this work we will adopt a further modification of the
reweighting factor, that has the advantage of yielding a Higgs
rapidity distribution that coincides exactly with the NNLO result: 
\begin{equation}
\mathcal{W}\left(y,\, p_{{\scriptscriptstyle \mathrm{T}}}\right)=h\left(\pt\right)\,\frac{\smallint d\sigma^{{\scriptscriptstyle \mathrm{NNLO\phantom{i}}}}\,\delta\left(y-y\left(\Phi\right)\right)-\smallint d\sigma_{B}^{{\scriptscriptstyle \mathrm{MINLO}}}\,\delta\left(y-y\left(\Phi\right)\right)}{\smallint d\sigma_{A}^{{\scriptscriptstyle \mathrm{MINLO}}}\,\delta\left(y-y\left(\Phi\right)\right)}+\left(1-h\left(\pt\right)\right)\,,\label{eq:NNLOPS-overall-rwgt-factor-1}
\end{equation}
which leads precisely to 
\begin{eqnarray}
\left(\frac{d\sigma}{dy}\right)^{{\scriptscriptstyle \mathrm{NNLOPS}}} & = & \left(\frac{d\sigma}{dy}\right)^{{\scriptscriptstyle \mathrm{NNLO}}}\,.\label{eq:NNLOPS-NNLOPS-eq-NNLO_0+MINLO_1-1}
\end{eqnarray}

The purpose of the $h$ profile function is quite similar to
what is done sometimes in \POWHEG{}, when the real emission cross
section is separated into a singular and a finite part~\cite{Nason:2004rx,Alioli:2008tz,Nason:2012pr}.
The only difference, in this case, is that, rather than an inclusive
LO-to-NLO correction, here we include an NLO-to-NNLO correction. This correction,
in the fixed order calculation, is concentrated in the region of zero
transverse momenta of the radiated partons, while in a resummed calculation
like \NLOPS{} or \NNLOPS{}, this is no longer the case, the zero
transverse momentum region being suppressed by a Sudakov form factor.
Thus, the correction must be spread over a range of non-zero transverse
momentum.

To facilitate a more intuitive understanding, we point out that in
the limit $\hgam\rightarrow\infty$, $h\left(p_{{\scriptscriptstyle \mathrm{T}}}\right)\rightarrow\theta\left(\hc\, m_{{\scriptscriptstyle \mathrm{H}}}-p_{{\scriptscriptstyle \mathrm{T}}}\right)$.
Thus, taking for example the leading jet transverse momentum to define
the argument $p_{{\scriptscriptstyle \mathrm{T}}}$ of the $h$ function,
we see that $d\sigma_{A}$ and $d\sigma_{B}$ in eqs.~(\ref{eq:NNLOPS-dsig-eq-dsig0-plus-dsig1}-\ref{eq:NNLOPS-dsig1})
are nothing more than the usual 0- and $\ge1$-jet cross sections.
Hence, in this limit, the reweighting procedure merely amounts to
rescaling the weights of the 0-jet events by the ratio of their respective
NNLO-to-NLO cross sections, albeit differentially in the Higgs boson's
rapidity. More generally, moving away from $\hgam=\infty$ towards
finite values, the effect on $h\left(p_{{\scriptscriptstyle \mathrm{T}}}\right)$
is to smear 
the step in the $\theta\left(\hc\, m_{{\scriptscriptstyle \mathrm{H}}}-p_{{\scriptscriptstyle \mathrm{T}}}\right)$
function. The $h$ profile function is therefore most easily
thought of as a smeared step function. In this work we have only performed
studies with $\hgam=2$. Nevertheless we consider the results we find
with this parameter choice to be wholly satisfactory.

Turning to the $\hc$ parameter in the profile function, one sees
that increasing $\hc$ increases the $d\sigma_{A}$ component of the
cross section relative to $d\sigma_{B}$. In fact, in the limit $\hc\to\infty$,
$d\sigma_{B}=0$ (eq.~\ref{eq:NNLOPS-dsig1}) and one recovers the
simple global event reweighting of eqs.~(\ref{eq:NNLOPS-proof-W-is-1-plus-aS2-i}-\ref{eq:NNLOPS-proof-W-is-1-plus-aS2-iii}).
If we choose $\hc\approx1$, the NNLO correction factor in $\mathcal{W}\left(y,\, p_{{\scriptscriptstyle \mathrm{T}}}\right)$
$ $is applied in a region where radiation is not much harder than
$m_{{\scriptscriptstyle \mathrm{H}}}$. It is also clear that we cannot
take $\hc\ll1$; if we do so, the NNLO correction will be concentrated
in a small region of the radiative phase space, $\pt\lesssim\hc\, m_{{\scriptscriptstyle \mathrm{H}}}\ll\mh$.
It thus becomes a delta function as $\hc\to0$, spoiling the accuracy
of the resummation.  
Effectively, $\hc$ must be of order one to avoid such a pathology.

The $\hc$ parameter shares some features with the ratio
of the resummation scale to the heavy boson mass in matched NNLO analytic
resummation calculations (see e.g.~\cite{Dasgupta:2001eq}).
In conventional resummation calculations the resummation scale affects
both the logarithms which are resummed and also how far
\emph{both} the hard NLO and NNLO virtual corrections are to be
distributed along the transverse momentum spectrum. Here $\hc$ plays 
an analogous role but it only affects the distribution of the hard
NNLO virtual corrections; the argument of the logarithms being resummed
and the distribution of NLO virtual corrections is unaffected by it. 
By the same analogy one should consider the `sensible' range in which
to vary $\hc$ as being limited to the same range in which the resummation
scale (divided by $m_{{\scriptscriptstyle \mathrm{H}}}$) is varied
in conventional analytic resummation calculations.

Before ending our discussion on the $\hc$ parameter, we wish to emphasise
that while its precise value is a source of systematic uncertainty,
it is fundamentally different to the \emph{merging} \emph{scales}
encountered in all other recent attempts to merge \NLOPS{} simulations
for multi-jet processes; even if a NNLO reweighting of such simulations
were to be admissible, e.g. by somehow having the equivalent NNLO
reweighting function of the form \mbox{$1+\mathcal{O}\left(\alpha_{{\scriptscriptstyle \mathrm{S}}}^{2}\right)$},
the merging scale would remain as an additional source of systematic
uncertainty. 
Moreover, while the dependence on $\hc$ is formally ${\cal O}(\as^5)$
or even zero in the case of inclusive quantities, in all other recent
\NLOPS{} merging attempts, the dependence on the respective merging
scales is ${\cal O}(\as^4)$ or worse.

Lastly, stepping back from the technicalities of the profile function
$h$, we point out that in these variant reweighting schemes
the extension of the proof of NNLO accuracy to variables other than
the inclusive Higgs boson rapidity spectrum follows that given earlier
with only trivial adjustments.

\section{Estimating uncertainties\label{sec:Estimating-uncertainties}}

We now examine the source of uncertainties in our NNLO generator,
and set up a method for determining its theoretical errors.

The uncertainties in the \HJMINLO{} generator are explored according
to the prescription we gave in ref.~\cite{Hamilton:2012np}.
There we have considered a 7-point scale variation, where all renormalization
scales appearing in the \MINLO{} procedure are multiplied by a scale
factor $K_{{\scriptscriptstyle \mathrm{R}}}$, and the factorization
scale is multiplied by a factor $K_{{\scriptscriptstyle \mathrm{F}}}$,
with 
\begin{equation}
(K_{{\scriptscriptstyle \mathrm{R}}},K_{{\scriptscriptstyle \mathrm{F}}})=(0.5,0.5),(1,0.5),(0.5,1),(1,1),(2,1),(1,2),(2,2)\,.\label{eq:Errors-KR-KF-list}
\end{equation}
We will consider the variation in our results induced by the above
procedure.

We compute $d\sigma^{\mathrm{NNLO}}/dy$ using the \HNNLO{} program
of ref.~\cite{Catani:2007vq,Grazzini:2008tf}. The theoretical uncertainty
in this calculation can be estimated by performing a factorization
and renormalization scale variation in the usual way. The choice of
the central scale has been subject of some debate. The value $m_{{\scriptscriptstyle \mathrm{H}}}$
has been used for a long time, but recently the value $m_{{\scriptscriptstyle \mathrm{H}}}/2$
seems to be preferred, on the grounds that it yields smaller NLO and
NNLO corrections. We will thus take $m_{{\scriptscriptstyle \mathrm{H}}}/2$
as the central scale choice for the NNLO calculation. Also in the
case of the NNLO calculation, we consider the variations of the renormalization
and factorization scales of eq.~(\ref{eq:Errors-KR-KF-list}), this
time applied at the central value $m_{{\scriptscriptstyle \mathrm{H}}}/2$.
This yields 49 variations in the \NNLOPS{} result. On the other
hand, we found that by limiting ourself to $K_{{\scriptscriptstyle \mathrm{R}}}=K_{{\scriptscriptstyle \mathrm{F}}}$
in the NNLO result no appreciable reduction of the scale variation
envelope is observed. We therefore restrict ourselves to this case,
thus ending up with 21 scale variation points. On top of this, we
have freedom in the choice of the $\pt$ variable and of the constant
$\hc$ of eq.~(\ref{eq:NNLOPS-hpT-fn-defn}). In the implementation we
have chosen to define the argument of the profile function, $\pt$,
as the transverse momentum of the hardest jet,
computed according to the inclusive $\kt$-algorithm with $R=0.7$.
This choice has the advantage that the region $\pt\to0$ is approached
only when all radiated partons have vanishing transverse momenta.
We have verified that other choices, such as the transverse momentum
of the Higgs, do not lead to significant differences. We assume as default
$\hc=0.5$, and consider variations between $0.5$ and $\infty$.

In order to perform our study, we have generated a single sample
of \HJMINLO{} events. The seven scale variation combinations have
been obtained by using the reweighting feature of the \POWHEGBOX{}.%
\footnote{ This feature is already optionally available in the current version
of the \POWHEGBOX{}, and will become a default feature in the upcoming
version 2.%
} The integrals $d\sigma_{A/B}^{\mathrm{MiNLO}}/dy$, needed in eq.~(\ref{eq:NNLOPS-overall-rwgt-factor-1}),
were performed and tabulated for each scale variation combination
using the \HJMINLO{} generated sample. Similarly, $d\sigma^{\mathrm{HNNLO}}/dy$
was tabulated for each of the three scale variation points. The analysis
is then performed by generating the \MINLO{} event with given values
of $(\Kr,\Kf)$, and multiplying its weight with the factor 
\begin{equation}
h\left(\pt\right)\times\,\frac{\smallint d\sigma_{(\Kr',\Kf')}^{{\scriptscriptstyle \mathrm{NNLO\phantom{i}}}}\,\delta\left(y-y\left(\Phi\right)\right)-\smallint d\sigma_{B,(\Kr,\Kf)}^{{\scriptscriptstyle \mathrm{MINLO}}}\,\delta\left(y-y\left(\Phi\right)\right)}{\smallint d\sigma_{A,(\Kr,\Kf)}^{{\scriptscriptstyle \mathrm{MINLO}}}\,\delta\left(y-y\left(\Phi\right)\right)}+\left(1-h\left(\pt\right)\right)\,.\label{eq:master}
\end{equation}
The central value is obtained by setting $(\Kr,\Kf)$ and $(\Kr',\Kf')$
equal to one, while to obtain the uncertainty band we apply this formula
for all the seven $(K_{{\scriptscriptstyle \mathrm{R}}},K_{{\scriptscriptstyle \mathrm{F}}})$
and three $(K_{{\scriptscriptstyle \mathrm{R}}}^{\prime},K_{{\scriptscriptstyle \mathrm{F}}}^{\prime})$
choices.

The conservative rationale/ansatz adopted here, in estimating errors
by varying the scales in the NNLO and NLO inputs in a fully independent way,
is essentially that we regard the uncertainties in the normalizations of
distributions, e.g. the transverse momentum spectrum of the Higgs boson,
as being independent of the respective uncertainties in the shapes ---
at least in the region covered by the profile function, $h(\pt)$, i.e.
that which includes the low $\pt$ domain. The former are determined by
the \HNNLO{} program, while the latter are due to the \HJMINLO{} input.
Outside of the low $\pt$ region, in the part corresponding to the
$1-h(\pt)$ term in eq.~(\ref{eq:master}), the uncertainty is given by
the standard \HJMINLO{} computation (which there corresponds to that of
conventional NLO with $\mur=\muf=\pt$ for the central scale choice).
Thus, in the low $\pt$ region the absolute uncertainty at a given point
in a distribution which is not inclusive, e.g. a $\pt$ spectrum, is
essentially given by the uncertainty in the shape at that point times
the uncertainty in the normalization. Readers familiar with such matters
may recognise the approach taken here as being analogous to the so-called
\emph{efficiency method}~\cite{Banfi:2012yh}, used for estimating errors
on cross sections in the presence of cuts; where uncertainties on the
theoretical predictions for the total cross section and the associated
efficiencies are assummed to be uncorrelated.

\section{Phenomenological analysis\label{sec:Results}}

In this section we present our phenomenological results. We consider
the production of a 125.5 GeV Higgs boson at the 8 TeV LHC. Throughout,
we use the MSTW8NNLO~\cite{Martin:2009iq} parton distribution functions
for all our results, including the \MINLO{} ones. We remark that
also the CT~\cite{Gao:2013xoa}, and NNPDF~\cite{Ball:2011uy} collaborations
have produced NNLO fits and could have been used in this context.
However, here we are not interested in PDF comparisons, and will stick
to a single set for simplicity.
The \NNLOPS{} events include parton
showering, as determined by \noun{Pythia 6}~\cite{Sjostrand:2003wg}, with Perugia
0 tune~\cite{Skands:2010ak} (\texttt{PYTUNE(320)}). We switched off hadronization
and multi-parton interactions, in order to carry out a more sensible
comparison with other parton level generators.

Throughout this paper, to define jets we used the anti-$k_T$
algorithm~\cite{Cacciari:2008gp} as implemented in
\FASTJET{}~\cite{Cacciari:2005hq,Cacciari:2011ma}.

As the reader may have noticed, there is a slight tension between the
choice of scale in the \HNNLO{} and \NNLOPS{} calculations, for observables
which are not fully inclusive with respect to all QCD radiation. In particular,
in \HNNLO{}, due to the nature of the calculation, it is not possible to use
a dynamical scale, as is done in \HJMINLO{}.\footnote{One could instead use
a regular NLO Higgs plus one jet calculation as implemented in \noun{Mcfm}
or \POWHEG{} with a dynamical scale, in comparing predictions for observables
in which the transverse momentum of the QCD radiation is non-zero.}
In this work we elect to use $\mh/2$ as the central scale in \HNNLO{}, as
input to our reweighting procedure, and in comparing to its predictions for
inclusive observables: since this setting is favoured by the community of
Higgs NNLO experts in determining the total inclusive cross section.
On the other hand, for jet cross sections at moderate and large transverse
momenta, that scale is generally considered to be too low.
Thus, when comparing to jet observables, we have instead used $\muf=\mur=\mh$
as the central scale choice in the \HNNLO{} \emph{predictions} (i.e. still
maintaining $\muf=\mur={\frac{1}{2}}\mh$ for the \HNNLO{} input to the
\NNLOPS{} reweighting procedure). In all cases, to obtain the \HNNLO{}
uncertainty band we perform a standard 7-point scale variation around the
central value.


\subsection{Higgs boson rapidity spectrum}

\begin{figure}[htb]
\begin{centering}
\includegraphics[clip,width=0.5\textwidth]{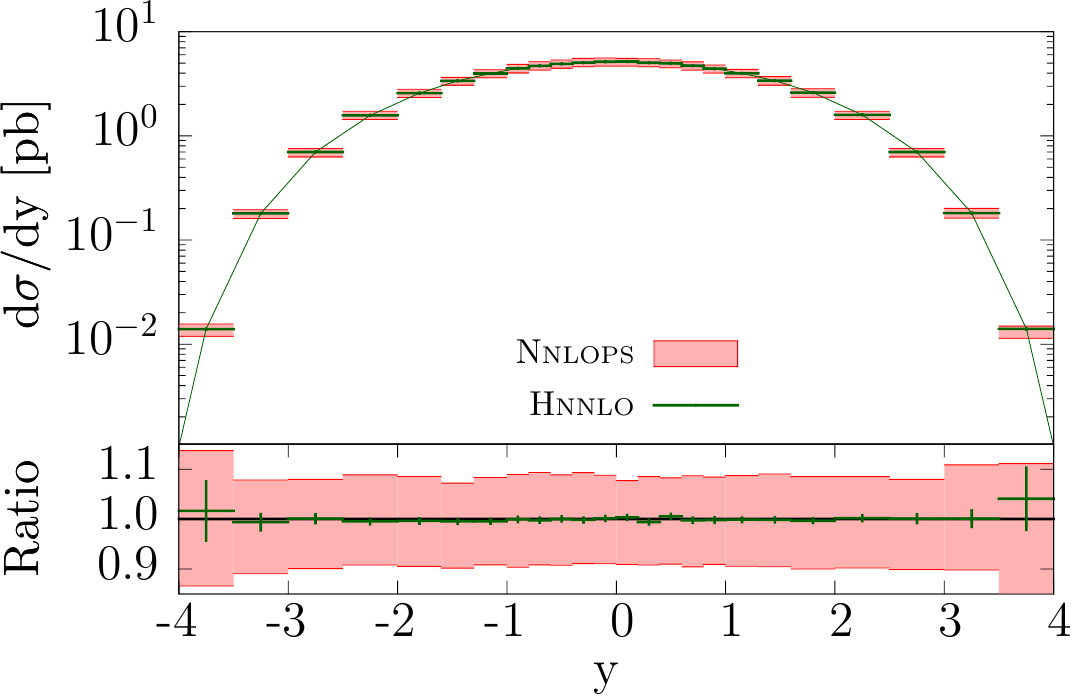}\hfill{}\includegraphics[clip,width=0.5\textwidth]{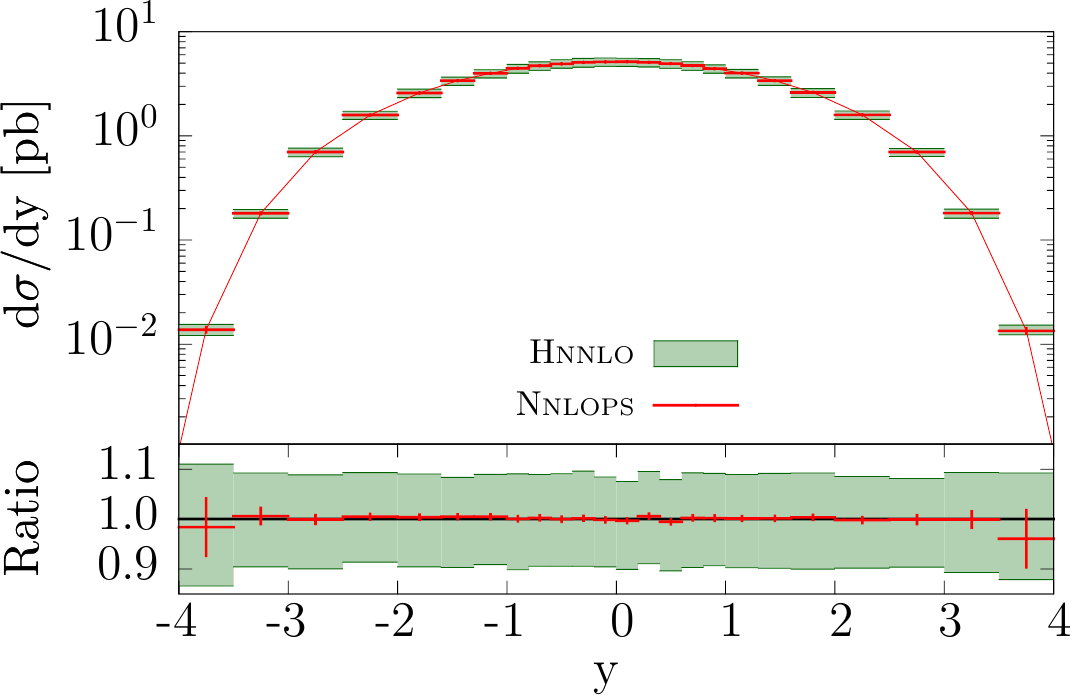} 
\par\end{centering}

\caption{Comparison of the \NNLOPS{} and \HNNLO{} results for the
  Higgs fully inclusive rapidity distribution. The \HNNLO{} central
  scale is $\muf=\mur=\mh/2$, and its error band is the 7-point scale
  variation envelope.  On the left (right) plot only the \NNLOPS{}
  (\HNNLO{}) uncertainty is displayed.  The lower left (right) panel
  shows the ratio with respect to the \NNLOPS{} (\HNNLO{}) prediction
  obtained with its central scale choice.}

\label{fig:yH-NNLOPS-band-vs-HNNLO-band} 
\end{figure}

We begin by showing in fig.~\ref{fig:yH-NNLOPS-band-vs-HNNLO-band}
the fully inclusive Higgs rapidity distribution. This, by construction,
should be identical in the \HNNLO{} and \NNLOPS{} calculations.

On the left, in the red shaded area, one can see the scale uncertainty
band predicted by the \NNLOPS{} simulation, with the conventional
fixed order \HNNLO{} result superimposed as green points. The lower
panel shows the ratio with respect to the \NNLOPS{} prediction obtained
with its central scale choice. On the right we have made the same
plots as on the left but with the \HNNLO{} predictions replacing
those of the \NNLOPS{} and vice versa; the scale uncertainty bands
are formed as described in Sec.~\ref{sec:Estimating-uncertainties}.
In the following we will compare the \NNLOPS{} to other results
with plots of the same kind. As expected, for this observable the
two calculations are in full agreement, both for their central values
and scale uncertainty envelopes; the latter being approximately $\pm$10\%
in size.

\subsection{Higgs boson transverse momentum}

\begin{figure}[htb]
\begin{centering}
\includegraphics[clip,width=0.65\textwidth]{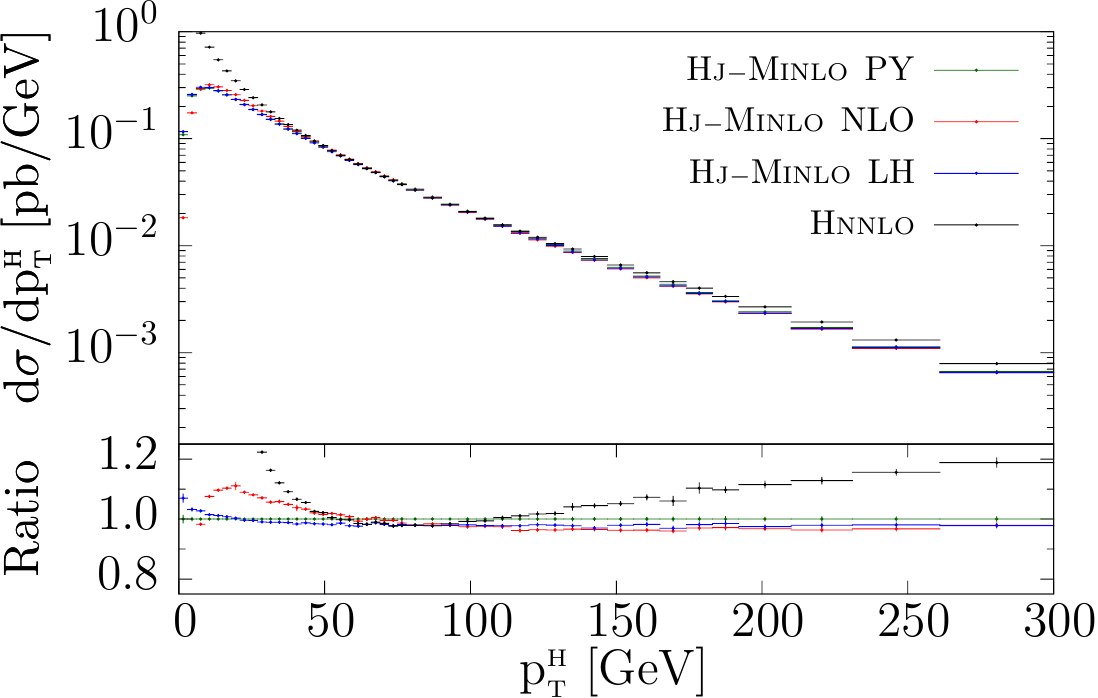} 
\par\end{centering}

\caption{Predictions for the Higgs boson transverse momentum spectrum: the
conventional NLO QCD prediction from \HNNLO{} with $\mu_{{\scriptscriptstyle \mathrm{R}}}=\mu_{{\scriptscriptstyle \mathrm{F}}}=m_{{\scriptscriptstyle \mathrm{H}}}$
(black), the \HJMINLO{} enhanced fixed order prediction (red), the \HJMINLO{}
result at the Les Houches event level (blue), and the \HJMINLO{}
result after showering (green). The lower panel shows the ratio relative
to the latter.}

\label{fig:pTH-HJ-MiNLO-NLO-vs-LH-vs-PY-vs-HNNLO} 
\end{figure}

Here, to begin with, we wish to discuss the evolution of the \NNLOPS{}
program's prediction, at each of the main stages of the simulation process,
as part of its validation and in order to provide relevant background,
before comparing it to state-of-the-art resummed calculations.
In figure~\ref{fig:pTH-HJ-MiNLO-NLO-vs-LH-vs-PY-vs-HNNLO} 
we show how the Higgs boson transverse momentum spectrum is
affected at the various phases of the event generation process in
the underlying \HJMINLO{} simulation (as described in sects.~2
and 3 of ref.~\cite{Hamilton:2012rf}): the \MINLO{} enhanced fixed
order prediction (red), the \HJMINLO{} hardest emission cross section
(blue) and the \HJMINLO{} events including parton shower effects
(green). The conventional NNLO QCD prediction from \HNNLO{} with
$\mu_{{\scriptscriptstyle \mathrm{R}}}=\mu_{{\scriptscriptstyle \mathrm{F}}}=m_{{\scriptscriptstyle \mathrm{H}}}$
is shown in black. In the lower panel all predictions in the upper
panel are shown as a ratio with respect to the central \HJMINLO{}+\PYTHIA{}
prediction. All of these predictions have the same $\mathcal{O}\left(\alpha_{{\scriptscriptstyle \mathrm{S}}}^{4}\right)$
accuracy if the small transverse momentum region is excluded.

The first most obvious feature is the difference between the various
\HJMINLO{} predictions and those of the conventional fixed order
program \HNNLO{} in the low $\pt$ region, with the latter exhibiting
unphysical divergent behaviour, and the former displaying, instead,
the anticipated, physical, Sudakov peak. In the high $p_{{\scriptscriptstyle \mathrm{T}}}$
tail region all of the predictions are in good agreement; we have
verified that they are within each other's scale uncertainties. Nevertheless,
the lower renormalization and factorization scales at high transverse
momentum in the \HNNLO{} result leads to a less steeply falling
cross-section in that region, as evident from the lower panel where
the ratio to the \HJMINLO{} showered result is displayed.

We remind the reader that for $p_{{\scriptscriptstyle \mathrm{T}}}>m_{{\scriptscriptstyle \mathrm{H}}}$
the \noun{Hj-Minlo }program reverts to $\mu_{{\scriptscriptstyle \mathrm{R}}}=\mu_{{\scriptscriptstyle \mathrm{F}}}=p_{{\scriptscriptstyle \mathrm{T}}}$
with the \noun{Minlo }Sudakov\noun{ }form factor also being equal
to one in this region.

Looking among the three \noun{Hj-Minlo }predictions, at high and
moderate $p_{{\scriptscriptstyle \mathrm{T}}}^{{\scriptscriptstyle
    \mathrm{H}}}$ the three distributions are almost indistinguishable
from one another, agreeing at the level of $\sim$2-3\%, very much
within the \noun{Hj-Minlo} scale uncertainty envelope --- a nearly
flat band, with a width of approximately $\pm20\%$ in the ratio
subplot (see fig.~3 of ref.~\cite{Hamilton:2012rf}).  This remarkable
agreement is easily understood as being due to the inclusivity of the
observable considered here (for more details see Appendix~\ref{app:A}).

Differences among the\noun{ Minlo} predictions only become apparent
in the low transverse momentum region below 50 GeV. We attribute these
more prominent deviations as being due to the amplification of 
NNLO sized differences (see Appendix~\ref{app:A}), between the hardest
emission cross section (blue) and enhanced fixed order (red), by large
logarithms of $p_{{\scriptscriptstyle \mathrm{T}}}^{{\scriptscriptstyle \mathrm{H}}}/m_{{\scriptscriptstyle \mathrm{H}}}$.
We also note the expected vanishing of the \noun{Hj-Minlo} fixed order
prediction (red) at low $p_{{\scriptscriptstyle \mathrm{T}}}^{{\scriptscriptstyle \mathrm{H}}}$,
by virtue of the fact that the NLO computation includes a Sudakov
form factor which is a function of the Higgs boson's transverse momentum.
The \noun{Hj-Minlo }hardest emission cross section and parton shower
level predictions smear out this region, by transforming underlying
Born configurations with low $p_{{\scriptscriptstyle \mathrm{T}}}^{{\scriptscriptstyle \mathrm{H}}}$
into configurations where the Higgs boson has zero transverse momentum,
through generating the hardest emission and also subsequent, shower
emissions in the latter case.

The differences in the region below 50 GeV can also be understood
from a different point of view, by noting that for the case that the
observable $O$ is the total inclusive cross section, it is clear
that the \noun{Hj-Minlo} enhanced fixed order prediction and its derivatives,
namely, the hardest emission cross section and subsequent parton showered
predictions, all agree identically; thus, a relatively low contribution
in the vicinity of the $p_{{\scriptscriptstyle \mathrm{T}}}^{{\scriptscriptstyle \mathrm{H}}}\sim0$
GeV in the case of the enhanced fixed order prediction (red) must
be compensated by it having a relatively high contribution elsewhere,
in this case the region $\sim15-50$ GeV.

In summary, we have seen that the \HJMINLO{}
predictions at the NLO, Les Houches, and showered level are
in close agreement, the largest discrepancy
(near $10\%$) occurring in the Sudakov region, where
effects beyond NLO are numerically significant, for reasons which
are well understood.

\begin{figure}[htb]
\begin{centering}
\includegraphics[clip,width=0.65\textwidth]{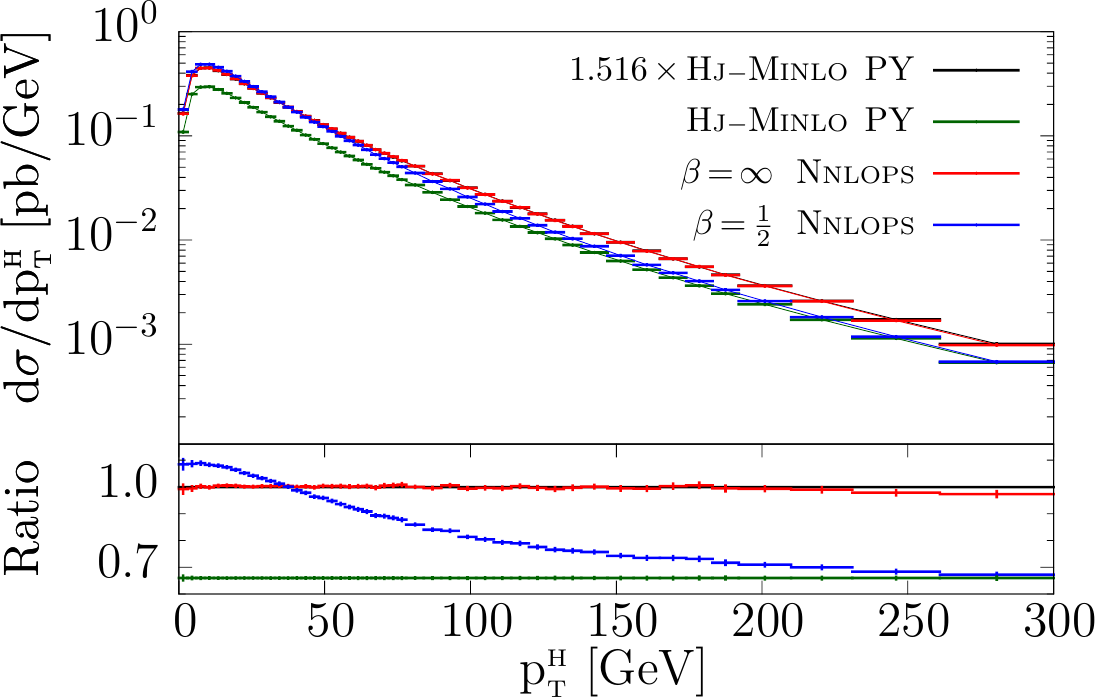} 
\par\end{centering}

\caption{Transverse momentum spectrum of the Higgs boson from the
\NNLOPS{} simulation with  $\hc=\infty$ (red) and our default $\hc=\frac{1}{2}$
(blue), compared to the \HJMINLO{} output (green).
The \HJMINLO{} output (rescaled by
a global factor such that the total inclusive cross section is 
the same as for the two \NNLOPS{} predictions)
is shown in black, and is almost exactly under the red line.
The ratio plots
are normalized to the black line.}

\label{fig:pTH-NNLOPS-vs-MiNLO} 
\end{figure}

 The
final ingredient to reach the NNLO accuracy is the inclusion of the
reweighting procedure discussed in sections~\ref{sub:NNLOPS} and
\ref{sub:Simple-variations}. In figure~\ref{fig:pTH-NNLOPS-vs-MiNLO}
we display the effect of the inclusion of the NNLO reweighting with
respect to the \HJMINLO{} result, for $\hc=\infty$ and $\hc=\frac{1}{2}$.
In the $\hc=\infty$ case, the NNLO reweighting can be well modeled
by an overall \emph{K}-factor, that does not modify the shape of the
transverse momentum distribution at all. This is easily understood, since in
practice the reweighting factor has a fairly mild dependence upon
the rapidity. By introducing a finite $\hc$ we do instead alter the
shape of the transverse momentum distribution, since, in this case,
the \emph{K}-factor is only applied to the lower portion of the $\pt$ spectrum.
We observe that the NNLO correction factor is quite large, around
1.5, in the small transverse momentum region, where the bulk of the
cross section lies. We remind the reader that in carrying out the
reweighting here, we have set $\muf=\mur={\frac{1}{2}}\mh$ in the
\HNNLO{} program and used the default \HJMINLO{} settings (which
correspond well, in the case of inclusive quantities, to conventional
NLO predictions with $\muf=\mur=\mh$). Had we chosen $\muf=\mur=\mh$
in determining the \HNNLO{} input to the reweighting procedure,
the correction factor would be near 1.3.

\begin{figure}[htb]
\begin{centering}
\includegraphics[clip,width=0.5\textwidth]{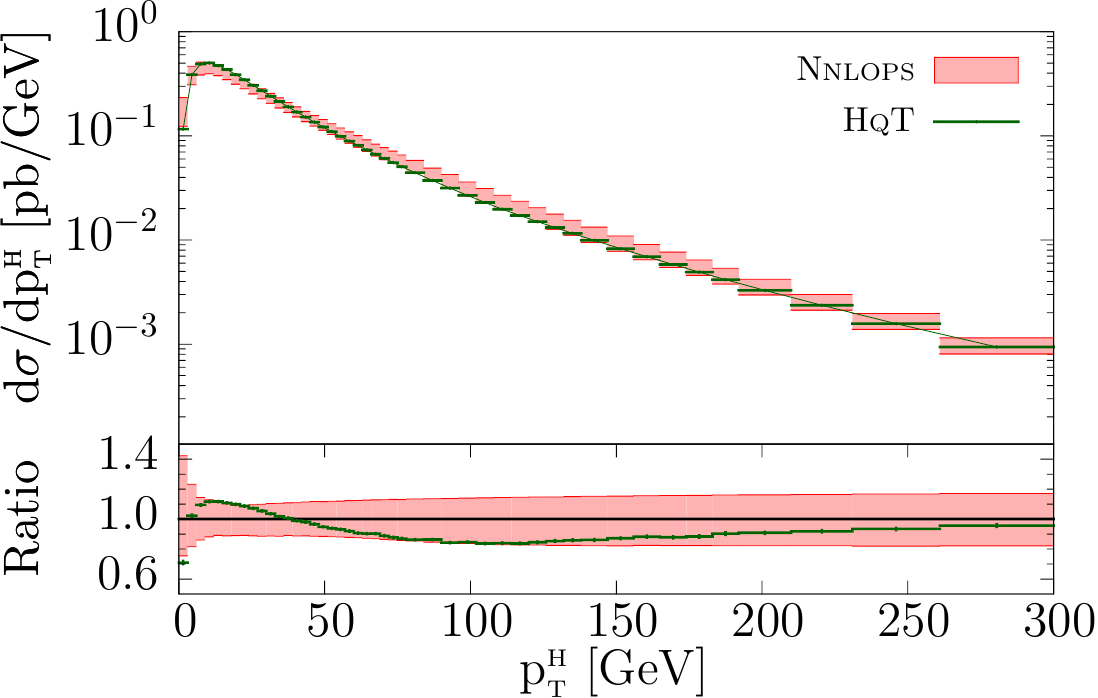}\hfill{}\includegraphics[clip,width=0.5\textwidth]{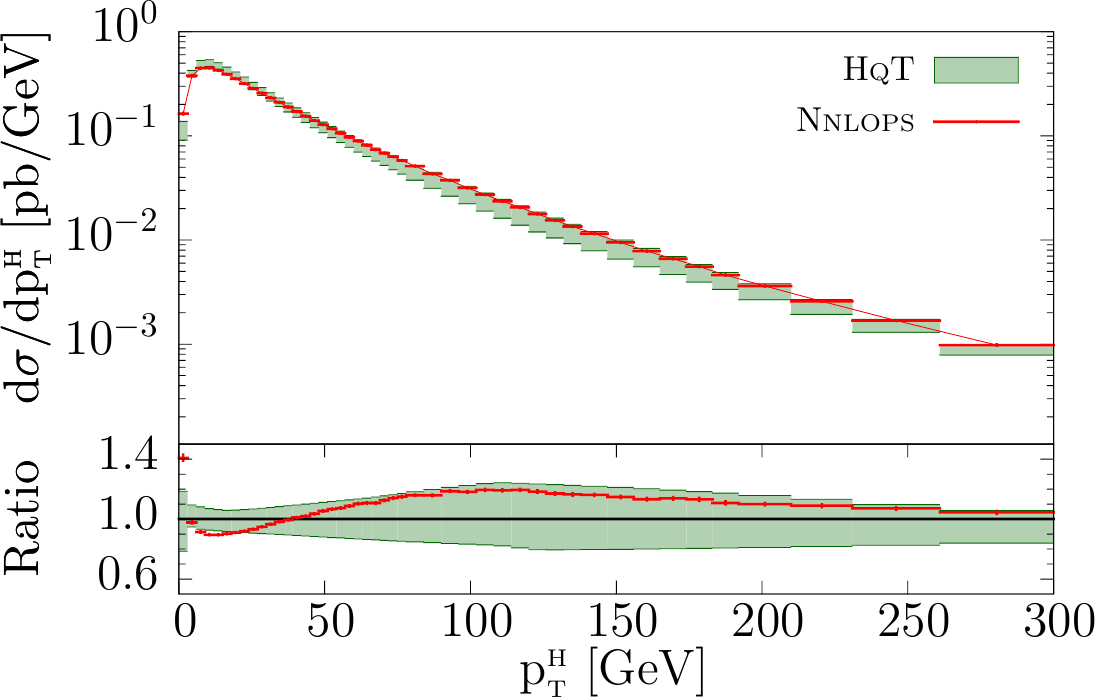} 
\par\end{centering}

\caption{Comparison of the $\hc=\infty$ \NNLOPS{} (red) with the NNLL+NNLO
prediction of \HQT{} (green) for the Higgs transverse momentum.
In \noun{HqT} we choose $\mur=\muf=\frac{1}{2}\mh$ as the central
scales, and keep the resummation scale always fixed to
$\frac{1}{2}\mh$.
On the left (right), the \NNLOPS{} (\HQT{}) uncertainty band is
shown. In the lower panel, the ratio to the \NNLOPS{} (\HQT{})
central prediction is displayed.}

\label{fig:pTH-NNLOPS-pTstar-0-vs-HqT} 
\end{figure}

\begin{figure}[htb]
\begin{centering}
\includegraphics[clip,width=0.5\textwidth]{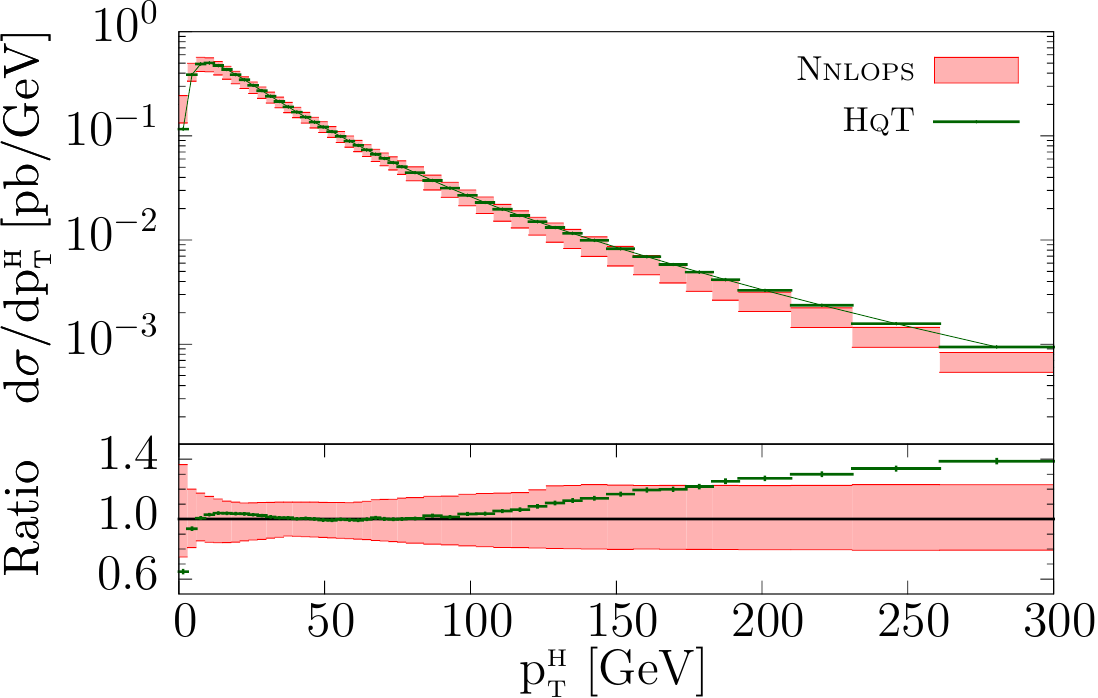}\hfill{}\includegraphics[clip,width=0.5\textwidth]{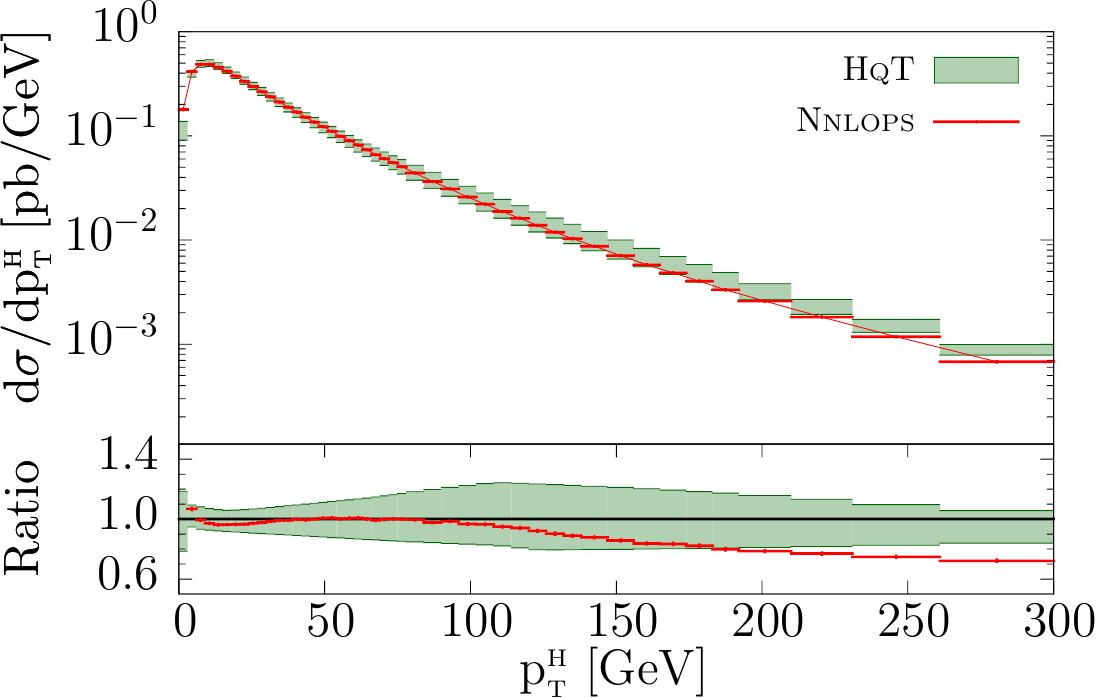} 
\par\end{centering}

\caption{As in fig.~\ref{fig:pTH-NNLOPS-pTstar-0-vs-HqT} but with 
$\hc=\frac{1}{2}$ in the profile function.}

\label{fig:pTH-NNLOPS-pTstar-default-vs-HqT} 
\end{figure}

In figures~\ref{fig:pTH-NNLOPS-pTstar-0-vs-HqT} and \ref{fig:pTH-NNLOPS-pTstar-default-vs-HqT},
we compare the \NNLOPS{} (see eq.~(\ref{eq:master})) with the \noun{HqT}
\cite{Bozzi:2005wk,deFlorian:2011xf} result for two choices of the
$\hc$ parameter in the profile function. The uncertainty band is
the envelope of the 21-point scale variation illustrated in section~\ref{sec:Estimating-uncertainties}.
We used the `switched' output of \noun{HqT}, forming the related uncertainty
band from the envelope of the seven results obtained by independent
variations of $\mur$ and $\muf$, by a factor of two, symmetrically,
about $\mur=\muf=\frac{1}{2}\mh$, while keeping the resummation scale
always fixed to $\frac{1}{2}\mh$.

Pleasingly, we see that the \noun{Nnlops }and \HQT{} results are
almost completely contained within each other's uncertainty band in
the region of moderate transverse momenta. We have verified that at
high transverse momentum the \HQT{} prediction agrees identically
with that of \HNNLO{}, since the `switched' output in the former
uses the fixed order result in this region. It follows that here we
see the \HQT{} spectrum falling less rapidly than that of the \NNLOPS{}
simulation at large $\pth$. As was seen in fig.~\ref{fig:pTH-NNLOPS-vs-MiNLO}
and remarked upon in the related discussion, in the case of $\hc=\infty$,
the \NNLOPS{} result is very well approximated by that of \HJMINLO{}
multiplied by a uniform NNLO-to-NLO \emph{K}-factor of 1.5, leaving
the slope of the distribution unchanged. On the other hand, for $\hc=\frac{1}{2}$
(fig.~\ref{fig:pTH-NNLOPS-pTstar-default-vs-HqT}) the \emph{K}-factor
enhancement is predominantly concentrated in the region $\pt\lesssim\frac{1}{2}\mh$,
yielding a modest but marked change in the shape of the \NNLOPS{}
distribution. In this case, the discrepancy observed earlier, in fig.~\ref{fig:pTH-HJ-MiNLO-NLO-vs-LH-vs-PY-vs-HNNLO},
between the fixed order calculation (\HNNLO{}) and the \HJMINLO{}
result at high transverse momentum remains unaltered, both in terms
of the shape \emph{and} normalization of the spectrum in that region. 
Once the partial NNLO calculation of Higgs plus one
jet~\cite{Boughezal:2013uia} will be complete, a comparison to it will
certainly provide further insight. 

Notice that the scale uncertainty band in the \NNLOPS{} calculation, for
high Higgs $\pt$, is larger in the $\hc=1/2$ than in the $\hc=\infty$
case. This is easily understood. By reweighting we reduce the scale
dependence of the \HJMINLO{} result for the inclusive cross section,
so that at the end we have a scale variation that is appropriate
to the NNLO calculation. In other words, in the $\hc=\infty$ case,
reweighting also partially compensates the scale variation in the large
transverse momentum tail. This leads to an underestimate of the theoretical
error in this region, since the high transverse momentum tail, that
is only computed with NLO accuracy, gets a scale variation of relative
NNLO order. This is not the case for $\hc=1/2$, for which the associated
scale uncertainty is characteristic of the NLO accuracy at
high $\pt$.

We remark that the choice of the $\hc$ parameter is related, to some
extent, to the choice of the resummation scale in conventional resummed
calculations. In the latter, the hard function, i.e. the
factor $1+C_1 \as+\ldots$ in front of the resummed expression,
\footnote{See for example eq.~(5) in ref.~\cite{Catani:1992ua}.}
which embodies hard virtual corrections to the leading order process,
is present in the resummed component in the region below the
resummation scale $\pt<Q$.
Whereas, in conventional resummation calculations $Q$ governs the 
$\pt$ range for the full NNLO and NLO hard virtual corrections, as
well as affecting the argument of the logarithms being resummed, in the
\NNLOPS{} case the scale $\beta \mh$ only determines the extent of
the second order hard virtual corrections $\sim C_2 \as^{2}$.
Motivated by this correspondence, in this work we favour values
of $\hc$ not larger than one, the value
$\frac{1}{2}$ corresponding to the preferred choice of resummation
scale in \HQT{} ($Q=\frac{1}{2}\mh$). In the following we will thus
stick to the $\hc=\frac{1}{2}$ choice as our default. We remind the
reader, however, that in the $\hc=\infty$ case, the modification
of the spectrum at high transverse momentum is an effect of order
$\as^{5}$, i.e. beyond our intended accuracy.

In finishing this discussion we note that in the $\hc=\infty$ case
the level of agreement between \noun{HqT }and the \noun{Nnlops }is
better than between \noun{HqT }and the \noun{Nlops Powheg }Higgs production
program in fig.~22 of ref.~\cite{Dittmaier:2012vm}, while for the
optimal setting $\hc=\frac{1}{2}$ the agreement between \noun{HqT
}and the \noun{Nnlops }is quite satisfactory.

\subsection{Leading jet transverse momentum}

\begin{figure}[htb]
\begin{centering}
\includegraphics[clip,width=0.5\textwidth]{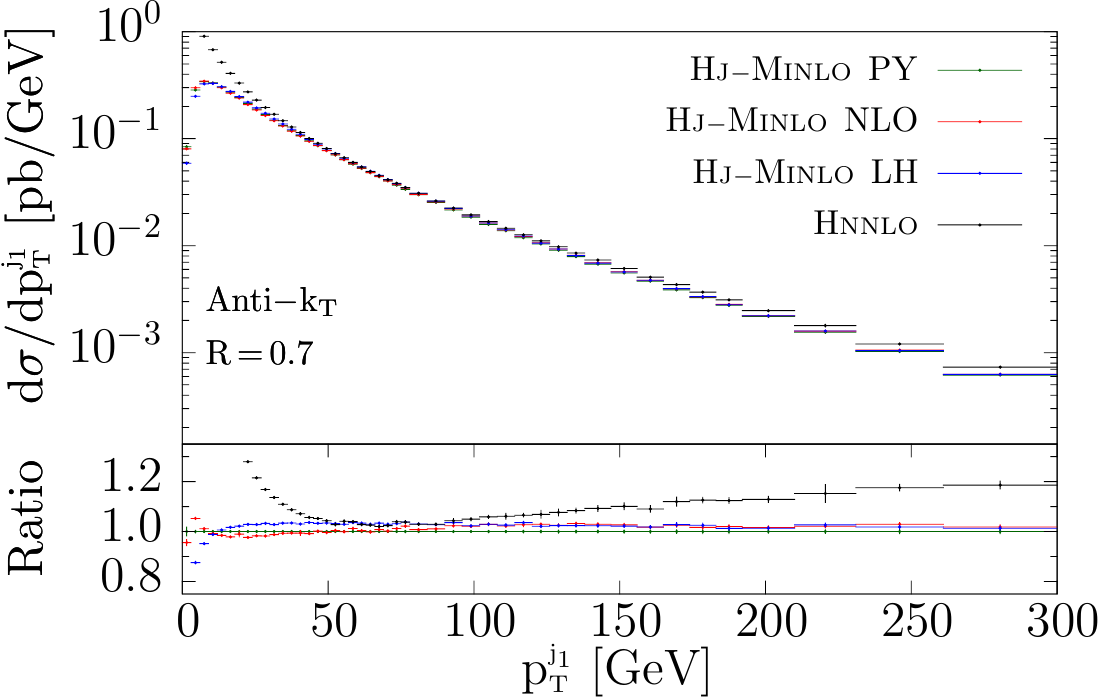}
\includegraphics[clip,width=0.5\textwidth]{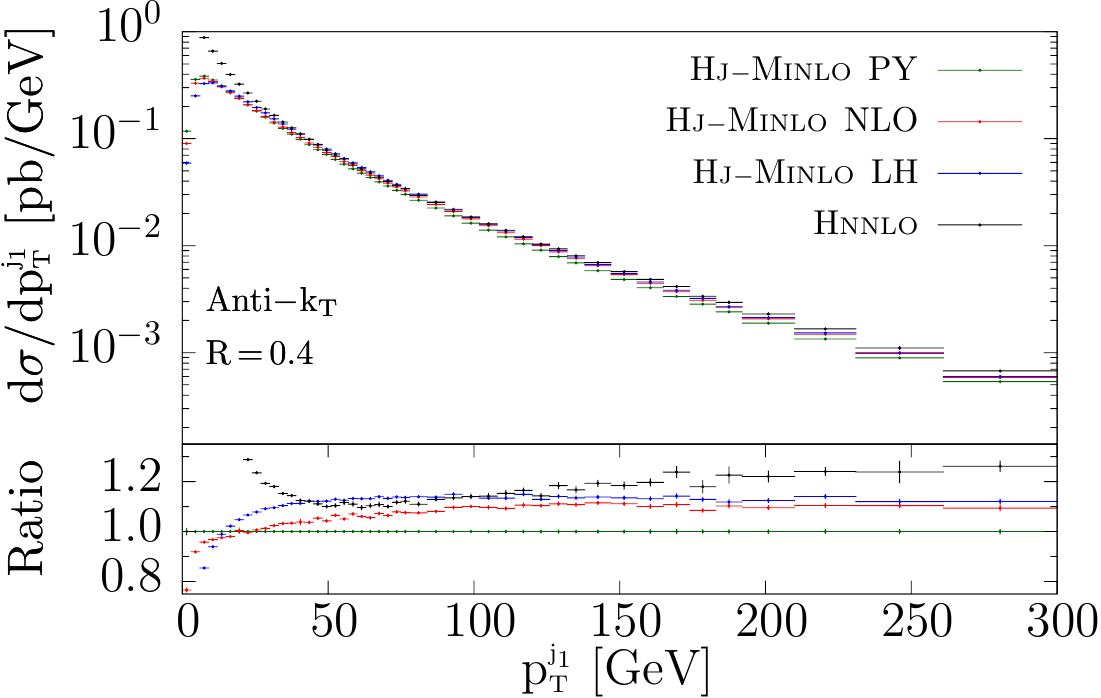} 
\par\end{centering}

\caption{Transverse momentum spectrum of the leading jet in Higgs boson production,
defined according to the anti-$\mathrm{k}_{{\scriptscriptstyle \mathrm{T}}}$-jet
algorithm, for radius $R=0.7$ (left) and
$R=0.4$ (right) jets. In each case we show the fixed order NNLO QCD
prediction from \HNNLO{} with $\mu_{{\scriptscriptstyle \mathrm{R}}}=\mu_{{\scriptscriptstyle \mathrm{F}}}=m_{{\scriptscriptstyle \mathrm{H}}}$
(black), the \HJMINLO{} enhanced fixed order prediction (red), the
\HJMINLO{} Les Houches event level (blue) and the \HJMINLO{} events
including parton shower effects (green). }

\label{fig:pTj1-HNNLO-v-MiNLO-vs-LH-vs-PY} 
\end{figure}

In this subsection we turn from the Higgs boson $\pt$ spectrum to
that of the leading jet, $\ptjone$. In figure~\ref{fig:pTj1-HNNLO-v-MiNLO-vs-LH-vs-PY}
we show results for the leading jet transverse momentum distribution,
all accurate up to and including $\mathcal{O}\left(\alpha_{{\scriptscriptstyle \mathrm{S}}}^{4}\right)$
contributions: the conventional NLO QCD prediction from \HNNLO{}
with $\mu_{{\scriptscriptstyle \mathrm{R}}}=\mu_{{\scriptscriptstyle \mathrm{F}}}=m_{{\scriptscriptstyle \mathrm{H}}}$
(black), the \HJMINLO{} enhanced fixed order prediction as described
in sects.~2 and 3 of ref.~\cite{Hamilton:2012rf} (red), the \HJMINLO{}
hardest emission cross section (\noun{Hj-Minlo LH}, blue), and the
ensuing parton shower level prediction from \PYTHIA{} (green).

In the case of $R=0.7$, in the left-hand plot, we see qualitatively
the same pattern of results as for the Higgs boson transverse momentum
spectrum. At least to leading order in perturbation theory, the jet
and the Higgs boson recoil against each other with equal and opposite
momenta in the transverse plane. At higher orders this picture is
modified, with multiple parton emissions leaking energy outside the
jet depleting its transverse momentum. We note that, for the cases
we consider, all $R$ dependence of the cross section originates from
real emission contributions to the \noun{Hj} process; more precisely,
for the cross section to have any $R$ dependence there must be at
least two partons in the final-state. Since the radius $R=0.7$ is
quite large, one expects any radiation leakage to be small, indeed,
in the limit $R\rightarrow\infty$ the jet algorithm clusters all
final state partons together into a single jet, which, by momentum
conservation, must exactly recoil against the Higgs boson. This being
the case we offer the same explanation for the pattern of results
shown in the left of fig.~\ref{fig:pTj1-HNNLO-v-MiNLO-vs-LH-vs-PY}
as in fig.~\ref{fig:pTH-HJ-MiNLO-NLO-vs-LH-vs-PY-vs-HNNLO}.

For $R=0.7$ jets we note just one small difference, namely, that
in the case of the Higgs boson transverse momentum spectrum, in the
vicinity of $\pt=0$ GeV, the enhanced fixed order prediction (red)
was greatly suppressed relative to the hardest emission cross section
and the showered result (blue and green) --- a deficit which was compensated
by a slight excess in the $\pt$ range 15-50 GeV --- while here it
is much more compatible with them, in fact, if anything there is a
slight excess of the former over the latter. This feature in the Higgs
boson transverse momentum spectrum was understood as being due to
the fact that the Sudakov form factor in the \noun{Minlo }formulation
of ref.~\cite{Hamilton:2012rf} is directly a function of the boson's
$\pt$, giving (formally) an infinite suppression at $\pt=0$ GeV.
In the present case, instead, a small $\pt$ of the hardest jet does
not imply the vanishing of the Higgs $\pt$ even at the NLO level,
and such suppression is not present, leading to better agreement in
the $\pt<50$~GeV region at each stage of the event generation process.

For radius $R=0.4$, the leading jet transverse momentum spectrum
is shown on the right of fig.~\ref{fig:pTj1-HNNLO-v-MiNLO-vs-LH-vs-PY}.
As in the case of $R=0.7$ at high transverse momentum the \HNNLO{}
result is less steeply falling than the \HJMINLO{} ones, due to the
different scale choice. 
We notice that for $R=0.4$ the \HJMINLO+\PYTHIA{} result is about 10\%
smaller than the \HJMINLO-LH one in the mid-to-high $\pt$ range.  In
fact, as pointed to in the preceding paragraph, multiple emission
parton shower effects cause the progenitor partons in the Les Houches
events to lose energy and hence reduce the cross sections for any jet
which had previously been associated to them. This leakage of
radiation outside the jet is clearly amplified for smaller jet radii.

\begin{figure}[htb]
\begin{centering}
\includegraphics[clip,width=0.5\textwidth]{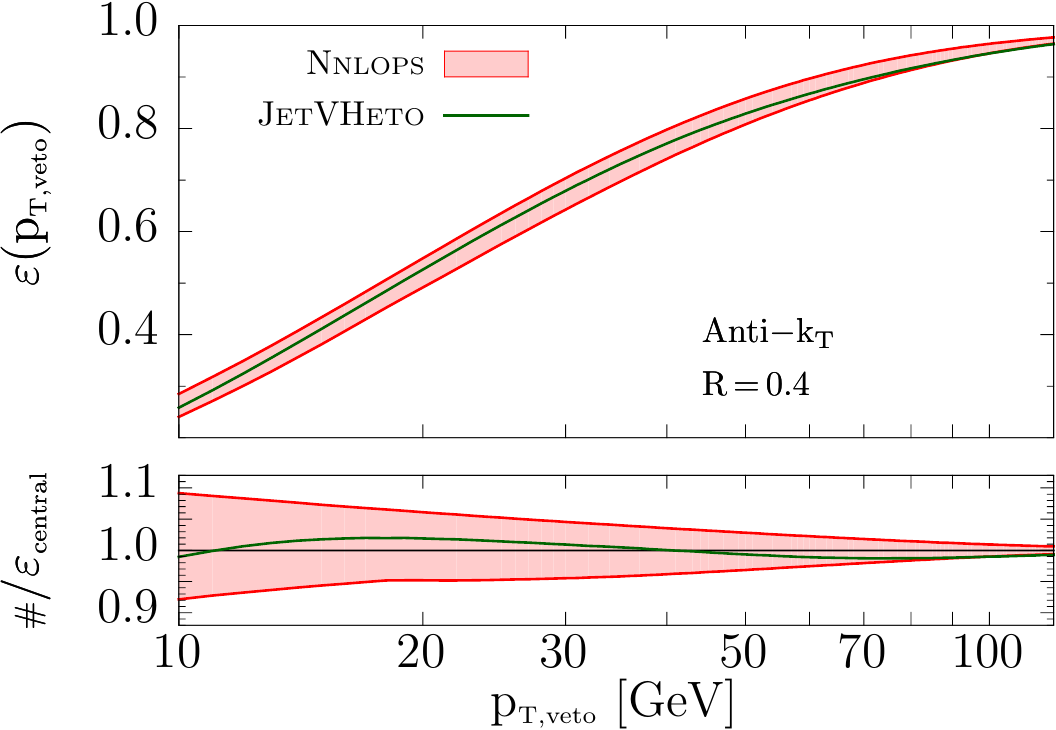}\hfill{}\includegraphics[clip,width=0.5\textwidth]{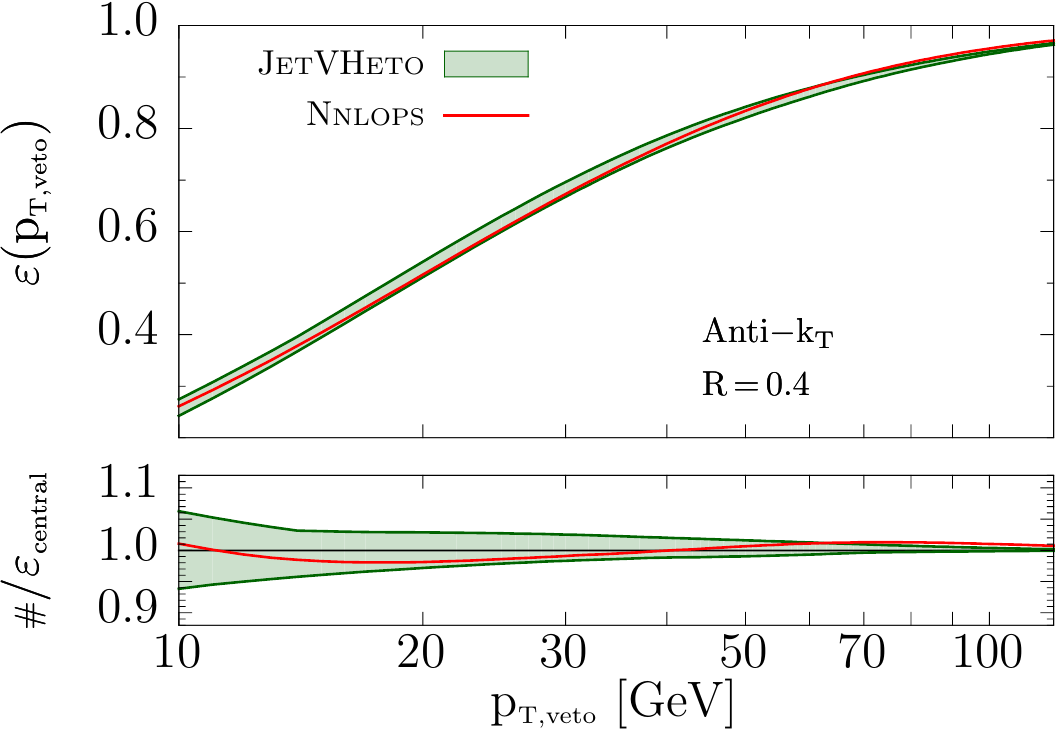} 
\par\end{centering}

\begin{centering}
\includegraphics[clip,width=0.5\textwidth]{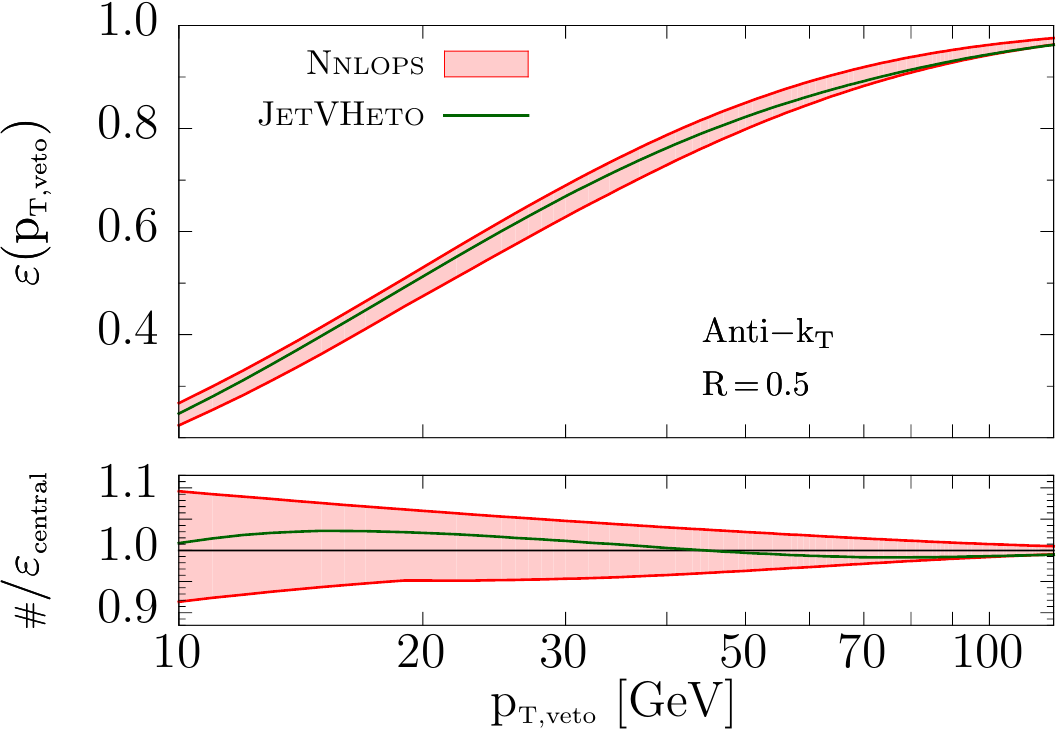}\hfill{}\includegraphics[clip,width=0.5\textwidth]{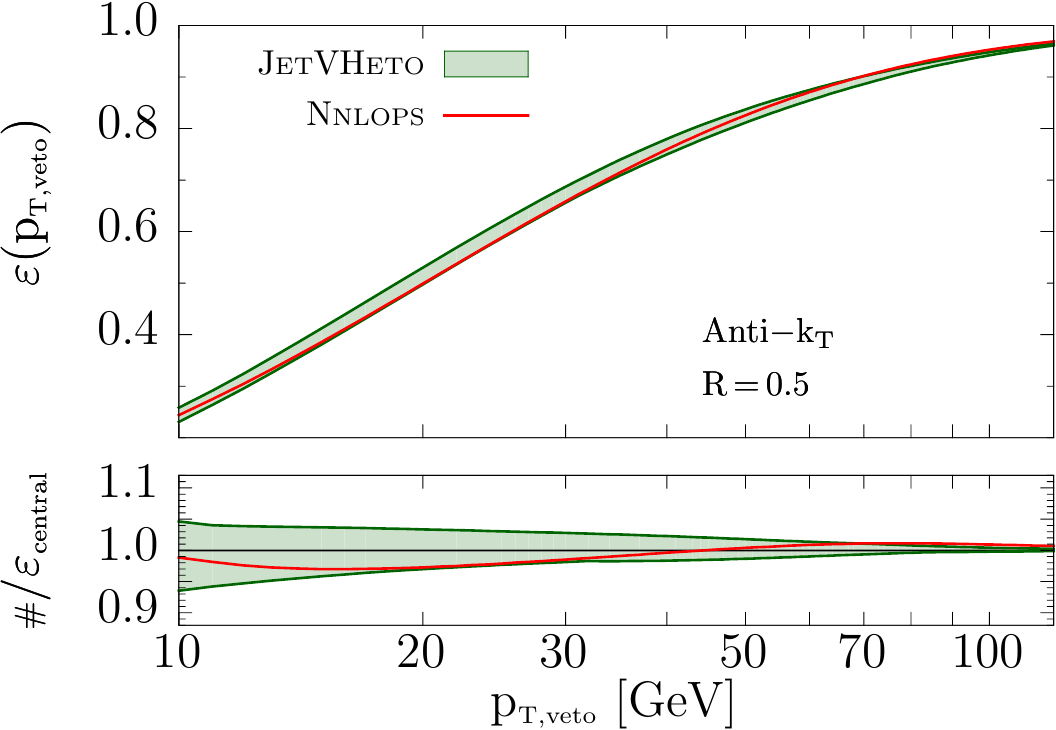} 
\par\end{centering}

\begin{centering}
\includegraphics[clip,width=0.5\textwidth]{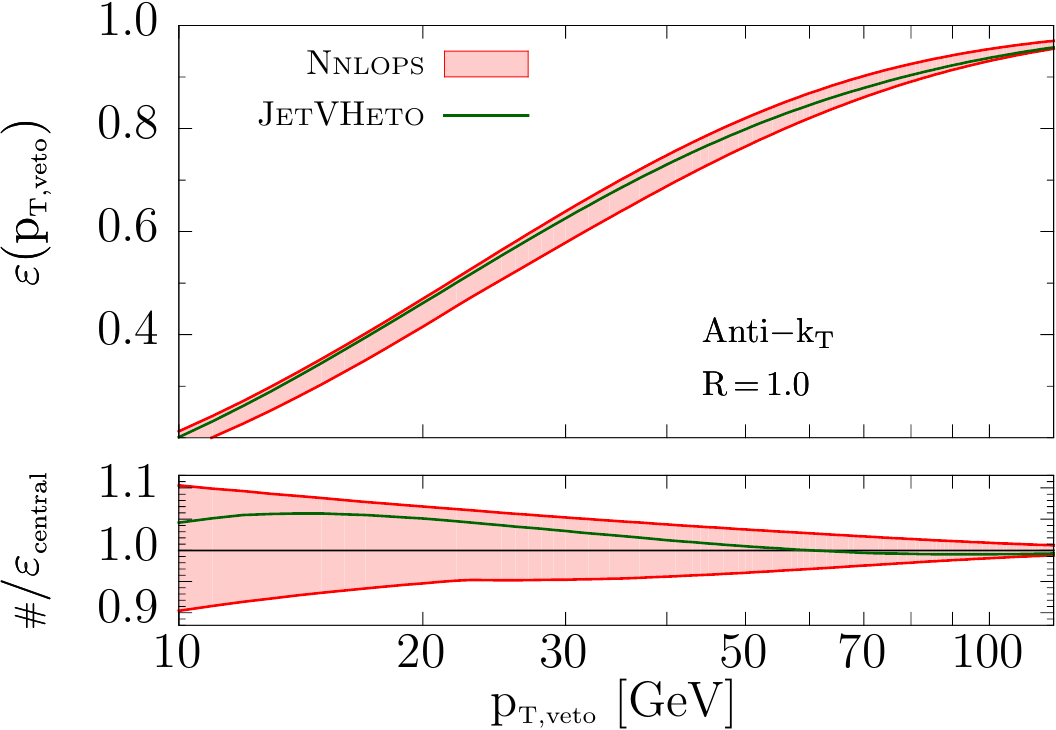}\hfill{}\includegraphics[clip,width=0.5\textwidth]{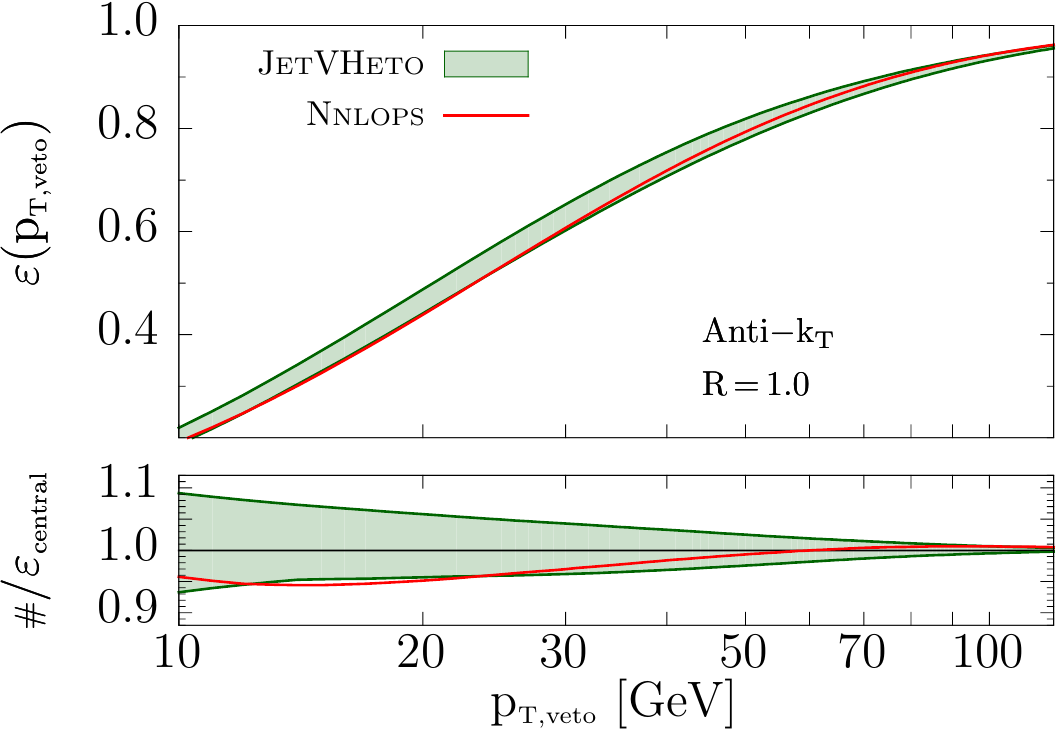} 
\par\end{centering}

\caption{The jet veto efficiency, $\varepsilon\left(p_{\scriptscriptstyle \mathrm{T,veto}}\right)$,
is defined as the cross section for Higgs boson production events
containing no jets with transverse momentum greater than $p_{\scriptscriptstyle \mathrm{T,veto}}$,
divided by the respective total inclusive cross section. In both plots
shown above we display the jet veto efficiency as a function of the
cut $p_{\scriptscriptstyle \mathrm{T,veto}}$. In the green shaded
area, one can see the scale uncertainty band obtained from the \NNLOPS{}
simulation (see Sect.~\ref{sec:Estimating-uncertainties} for details
regarding this uncertainty estimate), with the NNLL+NNLO uncertainty
band from the \JETVHETO{} program \cite{Banfi:2012jm,Banfi:2012yh}
superimposed in red. The lower pane displays the same quantities as
a ratio with respect to the central \NNLOPS{} prediction. The \NNLOPS{}
predictions here were obtained with the default profile function ($\hc=\frac{1}{2}$)
used in determining the NNLO reweighting $\mathcal{W}\left(y,\pt\right)$. }

\label{fig:pTj1-NNLOPS-vs-Jetvheto-bands} 
\end{figure}

In fig.~\ref{fig:pTj1-NNLOPS-vs-Jetvheto-bands} we have plotted
\NNLOPS{} and NNLL+NNLO \JETVHETO{} \cite{Banfi:2012jm} predictions
for the jet veto efficiency,
$\varepsilon\left(p_{\scriptscriptstyle \mathrm{T,veto}}\right)$,
defined as the cross section for Higgs boson production events
containing no jets with transverse momentum greater than
$p_{\scriptscriptstyle \mathrm{T,veto}}$, divided by the respective
total inclusive cross section,
\begin{equation}
\varepsilon\left(p_{\scriptscriptstyle \mathrm{T,veto}}\right) = 
\frac{1}{\sigma^{{\scriptscriptstyle \mathrm{tot}}}}
\int d\sigma\ \theta\left(p_{\scriptscriptstyle \mathrm{T,veto}}-
\pt^{{\scriptscriptstyle \mathrm{j_1}}}\right)\,.
\end{equation}
The jets considered here are formed
according to the anti-$k_{{\scriptscriptstyle \mathrm{T}}}$ jet
algorithm \cite{Cacciari:2008gp}, for a variety of different jet
radii; from top to bottom in
fig.~\ref{fig:pTj1-NNLOPS-vs-Jetvheto-bands} we have, pairwise,
$R=$0.4, 0.5, 1.0. In the left-hand column, in the red shaded area, we
show the scale uncertainty band predicted by the \NNLOPS{} simulation,
with the central NNLL+NNLO resummed prediction of \JETVHETO{}
superimposed in green (matching scheme-(a), $\mur=\muf=\mu_{{\scriptscriptstyle
    \mathrm{Q}}}=\frac{1}{2}\mh$, $\mu_{{\scriptscriptstyle
    \mathrm{Q}}}$ being the resummation scale).  The
lower panel shows the ratio with respect to the \NNLOPS{} prediction
obtained with its central scale choice. On the right we have made the
same plots as on the left but with the \JETVHETO{} predictions
replacing those of the \NNLOPS{} and vice versa.

The uncertainty band in the \JETVHETO{} results is the
envelope of a seven point variation of $\mur$ and $\muf$
by a factor of two. This is in contrast to the band associated with
it in the predictions of ref.~\cite{Banfi:2012jm}, where additionally
resummation scale and matching scheme variations were included in
the envelope. Thus the \JETVHETO{} error band here is considerably
smaller than that shown in ref.~\cite{Banfi:2012jm}. However, in
order to have a like-for-like comparison to the \NNLOPS{} band we
have restricted the \JETVHETO{} uncertainty estimate to the same
class of variations. Notwithstanding this more limited evaluation
of the uncertainties, we see the two sets of predictions still lie
within each other's error bands, except for a barely visible excursion
of the \NNLOPS{} outside the low edge of the \JETVHETO{} envelope
in the $R=1.0$ plot (fig.~\ref{fig:pTj1-NNLOPS-vs-Jetvheto-bands},
bottom-right). In all cases the central predictions of the \NNLOPS{}
and \JETVHETO{} programs are never out of agreement by more than
5-6\%, thus it becomes possible to meaningfully use the former in conjunction
with the more comprehensive, conservative, uncertainties of the latter
in real analysis.

\subsection{Next-to-leading jet transverse momentum }

\begin{figure}[htb]
\begin{centering}
\includegraphics[clip,width=0.5\textwidth]{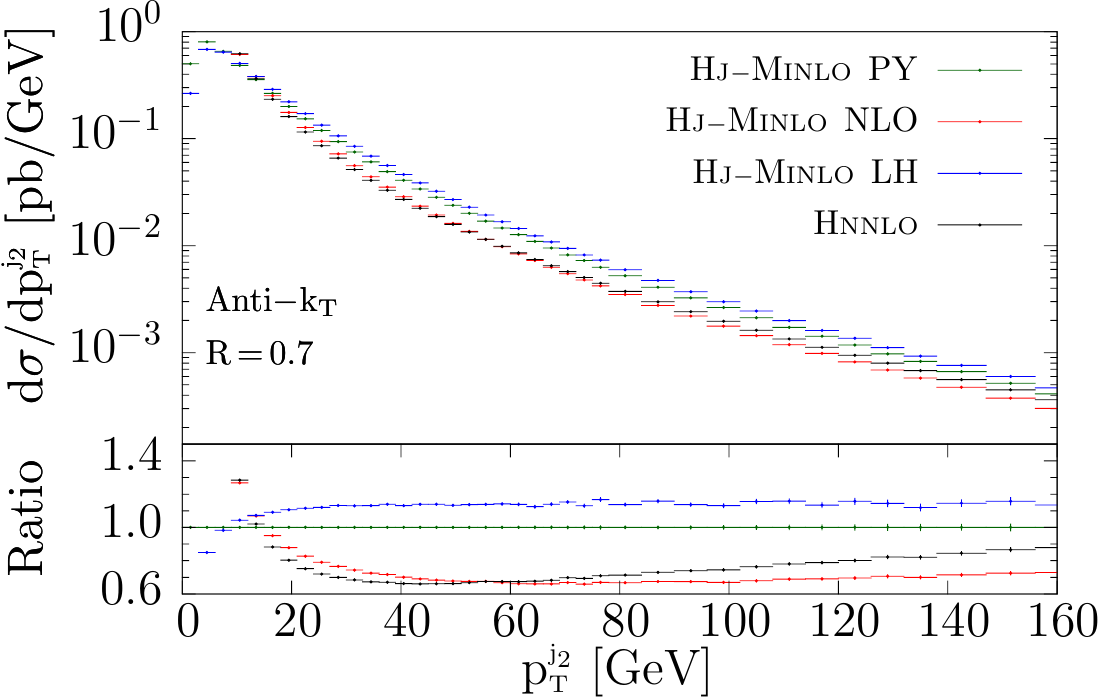}\hfill{}\includegraphics[clip,width=0.5\textwidth]{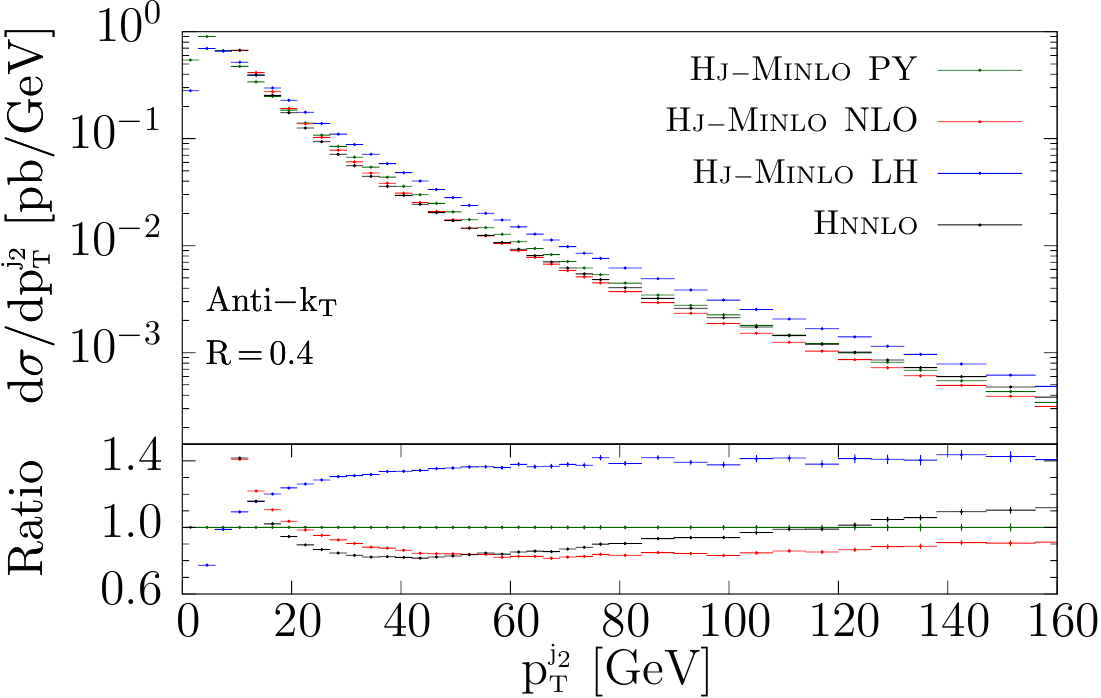} 
\par\end{centering}
\caption{Transverse momentum distribution of the second
leading jet in Higgs boson production. Jets are defined according to
the anti-$k_{{\scriptscriptstyle \mathrm{T}}}$ jet algorithm,
for $R=0.7$ (left) and $R=0.4$ (right). We show the
fixed order NNLO QCD predictions from \HNNLO{} with $\muf=\mur=\mh$ (black),
from \HJMINLO{}-NLO (red), from \HJMINLO{}
Les Houches event level (blue), and from \HJMINLO{} including parton
shower effects (green). }
\label{fig:pTj2-HNNLO-v-MiNLO-vs-LH-vs-PY} 
\end{figure}

We now briefly discuss the description of the second hardest jet
given by our \NNLOPS{} generator. First of all, we remind the reader
that our generator describes this distribution only at tree level
accuracy, but certainly introducing incomplete higher order corrections.
This is illustrated in fig.~\ref{fig:pTj2-HNNLO-v-MiNLO-vs-LH-vs-PY}
 where the second jet transverse
momentum distribution obtained with the \HNNLO{}, the \HJMINLO-NLO,
the \HJMINLO-LH and the fully showered \HJMINLO{} generators.
We see fair agreement between the \HNNLO{} and the \HJMINLO-NLO
predictions, as expected,
with previously discussed differences in scale assignments accounting for
discrepancies between the two in the high transverse momentum
tail. On the other hand, the
\HJMINLO-LH result is markedly higher than the fixed order results.
Since the second jet is certainly the \POWHEG{} hardest radiation
in this case, we identify the cause of this increase as due to the
fact that \POWHEG{} multiplies the hardest radiation spectrum by
its NLO \emph{K}-factor. Notice that in the case of $R=0.4$, subsequent
shower radiation leads to a transverse momentum spectrum that is
in better agreement with the fixed order calculation. This is due to
the fact that energy leakage outside the cone due to showering
softens the spectrum. However, the fact that this effect competes
with the \emph{K}-factor effect, up to the point of nearly canceling it,
has to be considered accidental. In fact, for $R=0.7$, where the
effect of energy leakage is negligible, we see no such compensation.
We also remind the reader that the so called \emph{K}-factor effect is
formally of order $\as^5$ or higher, beyond our intended accuracy.

\begin{figure}[htb]
\begin{centering}
\includegraphics[clip,width=0.45\textwidth]{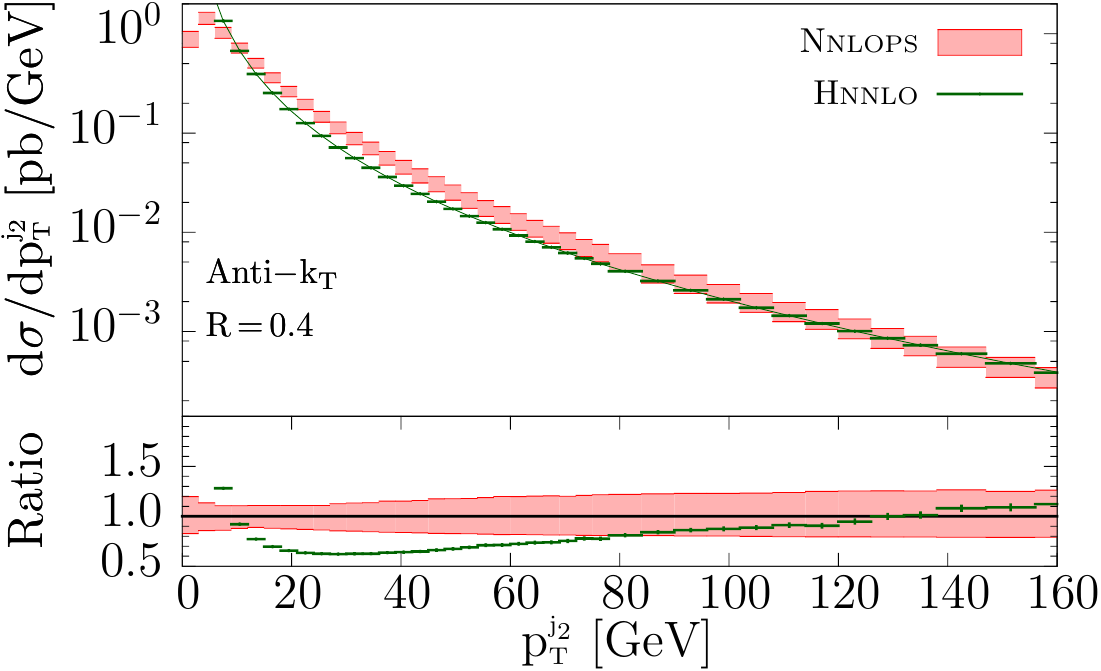}\hfill{}\includegraphics[clip,width=0.45\textwidth]{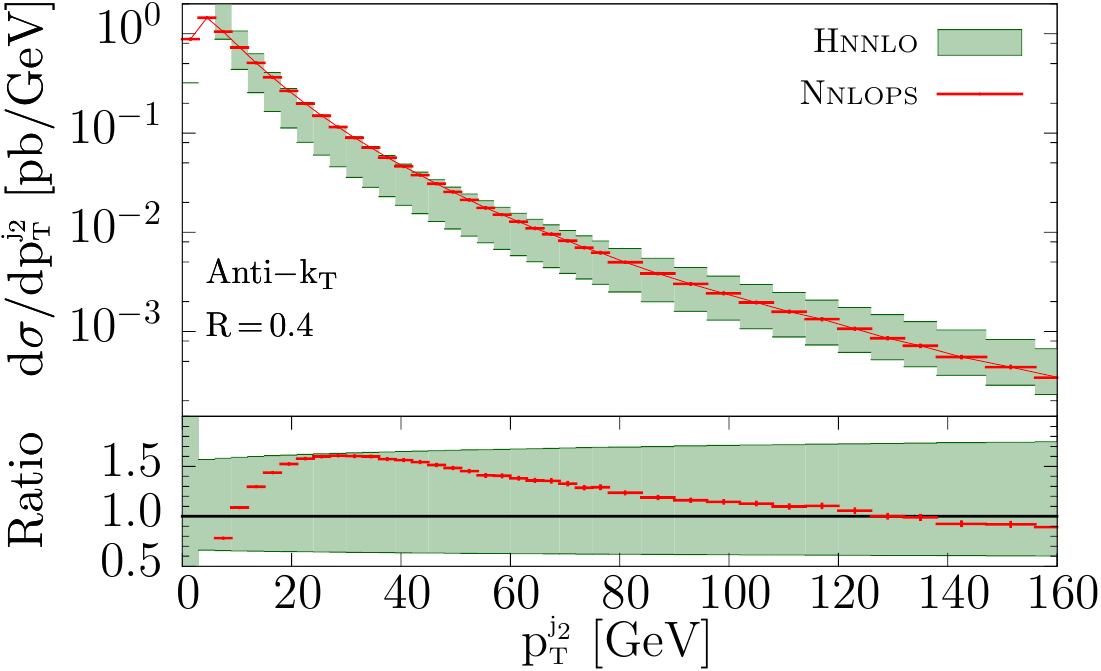} 
\par\end{centering}
\caption{Transverse momentum spectrum of the second leading jet
computed with the \NNLOPS{} and \HNNLO{} generators.
The \NNLOPS{} error band (displayed on the left plot) is obtained with our default method.
The \HNNLO{} band (on the right) is obtained with a 7-point scale variation
around the central value~$\muf=\mur=\mh$.}
\label{fig:pTj2-HNNLO-vs-NNLOPS-R-0.7} 
\end{figure}

\begin{figure}[htb]
\begin{centering}
\includegraphics[clip,width=0.5\textwidth]{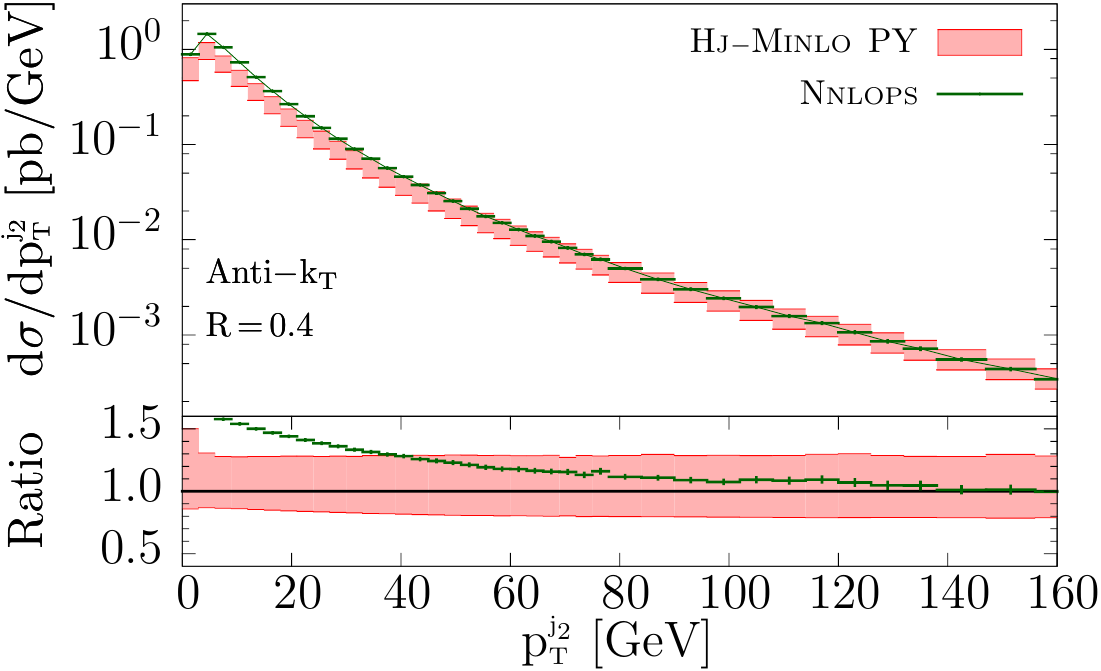}\hfill{}\includegraphics[clip,width=0.5\textwidth]{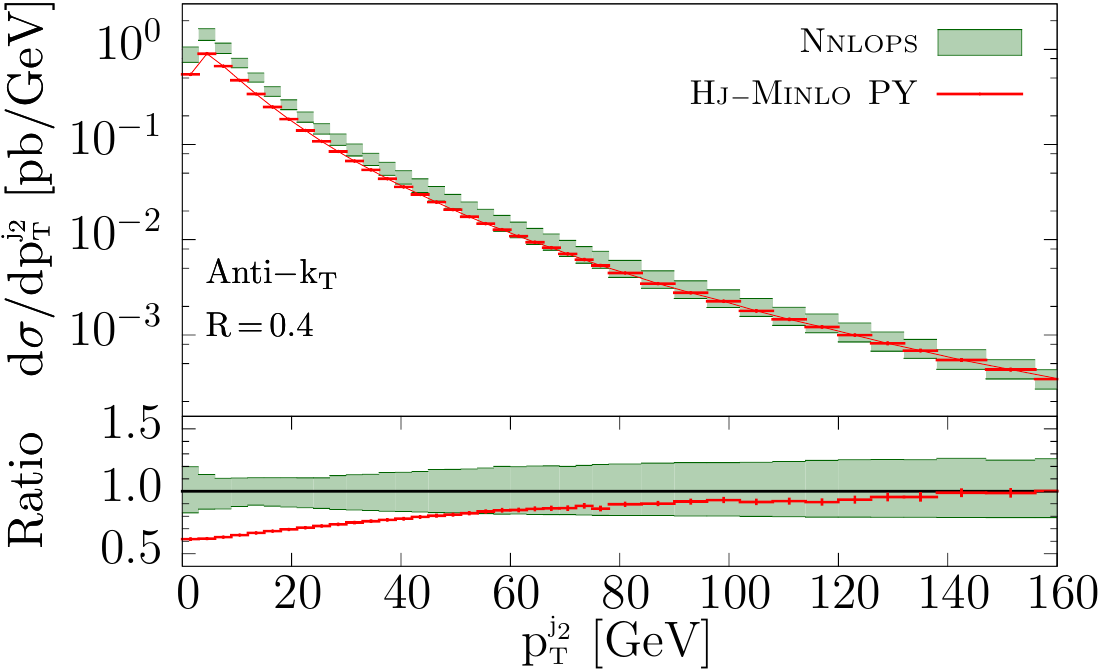} 
\par\end{centering}
\caption{Transverse momentum spectrum of the second leading jet
computed with the \HJMINLO{} and \NNLOPS{} generators.
The error band for the \HJMINLO{} generators (displayed in the left plot) are
obtained with the 7-point scale variation, while for error band of the \NNLOPS{}
calculation (on the right plot) we used
our default prescription.}
\label{fig:pTj2-HJ-MINLO-vs-NNLOPS-R-0.7} 
\end{figure}

The scale uncertainty in the radiation of the hardest jet is usually
underestimated in \POWHEG{}. In fact, the spectrum of the hardest
radiation is generated with a scale independent procedure. This
spectrum is multiplied by the underlying Born cross section, that
has a next-to-leading order scale dependence. Thus, also the \POWHEG{}
spectrum displays formally a next-to-leading order scale dependence.
In the case of the \NNLOPS{} generator, the reweighting procedure
constrains even more the underlying Born scale variation, forcing it
to integrate up to the NNLO scale dependence. This is apparent from
figs.~\ref{fig:pTj2-HNNLO-vs-NNLOPS-R-0.7}
and~\ref{fig:pTj2-HJ-MINLO-vs-NNLOPS-R-0.7}.
In fig.~\ref{fig:pTj2-HNNLO-vs-NNLOPS-R-0.7}
we compare the \NNLOPS{} and \HNNLO{} results for the second hardest jet
transverse momentum distribution. We use our default method to compute
the uncertainty band in the \NNLOPS{} result, while in the \HNNLO{}
case we present the standard 7-points scale variation taking $\muf=\mur=\mh$
for the central value.
We see that the \HNNLO{} envelope includes the \NNLOPS{} one, the latter
being considerably smaller.

In fig.~\ref{fig:pTj2-HJ-MINLO-vs-NNLOPS-R-0.7} 
we compare the \HJMINLO{} showered result with our \NNLOPS{} output. We see
that the main difference in the central value prediction is due to the fact
that the NNLO \emph{K}-factor is mainly applied when the transverse momentum of 
the hardest jet is below half the Higgs mass, and that we expect the
hardest jet transverse momentum to be just slightly above that
of the second jet. The two results also tend to lie outside of each other's
error band in the small transverse momentum region, a further indication
that the scale variation uncertainty is too small for this tree
level distribution. Furthermore, the \NNLOPS{} band is smaller than
the \HJMINLO{} one, as anticipated earlier.

\section{Conclusions\label{sec:Conclusions}}

In this paper we have presented a \NNLOPS{} implementation of Higgs boson
production in hadronic collisions, thus yielding the first example
of a NNLO calculation matched to a parton shower. We observe that our method
is the NNLO extension of the current \NLOPS{} methods. In fact,
in both \MCatNLO{} and \POWHEG{}, NLO accuracy is achieved without
the need of unphysical separation scales; NLO accuracy is granted
for infrared safe observables, and logarithmic accuracy is maintained
at the level of the shower Monte Carlo. These same features are achieved
at the NNLO level by our procedure.

Presently the method has been applied to Higgs production in the
large $m_{t}$ limit. The full inclusion of finite mass effects would
require a calculation of Higgs plus jet production at NLO including
finite mass effects, which is currently not available. It is possible,
however, to include such mass effects at least at $\mathcal{O}(\as^{3})$,
if we are allowed to treat them as small corrections. One could
incorporate such effects by reweighting the \HJMINLO{} events by the
ratio of the full $m_{b},m_{t}$
mass dependent Higgs+jet cross section at $\mathcal{O}(\as^{3})$
over its large $m_{t}$ limit, computed at the underlying Born
level. Then one could proceed as
in the present work, reweighting the events using the recent HNNLO-V2
calculation~\cite{Grazzini:2013mca},
which includes mass effects up to order $\mathcal{O}(\as^{3})$. We
leave this possibility to future studies.

Our procedure relies upon the results of ref.~\cite{Hamilton:2012rf},
in which a method was presented for extending an \NLOPS{} simulation
of Higgs plus jet production, so as to simultaneously deliver NLO accuracy
for inclusive quantities. At present, the method of ref.~\cite{Hamilton:2012rf}
has been applied, besides the Higgs case, to the Drell-Yan processes
and the Higgsstrahlung process~\cite{Luisoni:2013cuh}. It can be
generalized easily to all reactions involving the production of a heavy
colourless system. For more general processes, we see no obstacles to
the implementation of
the procedure of ref.~\cite{Hamilton:2012rf}, except for the increased
complexity in the calculation of the needed resummation formulae.
Thus, the present method for building \NNLOPS{} generator may be
extended to more complex processes, subject to the availability of
the corresponding NNLO calculation, and of the suitable extension
of the procedure of ref.~\cite{Hamilton:2012rf}. 

\section{Acknowledgements}
We gratefully acknowledge Carlo Oleari for important contributions
made earlier in the development of the \MINLO{} method. It is a
pleasure to thank Andrea Banfi, Massimiliano Grazzini and Gavin Salam
for helpful comments and discussions. GZ is supported by Science and
Technology Facility Council. 
We acknowledge ESI (ER and GZ) and KITP (GZ) for hospitality while part
of this work was carried out.

\appendix

\section{Inclusivity of the Higgs $\pt$ spectrum}
\label{app:A}

Following sect~4.3
of ref.~\cite{Frixione:2007vw}, suppressing indices for simplicity,
the difference between conventional NLO and \noun{Powheg }predictions
for an observable $O$, $\left\langle O\right\rangle ^{{\scriptscriptstyle \mathrm{PWG}}}=\left\langle O\right\rangle ^{{\scriptscriptstyle \mathrm{NLO}}}+\delta\left\langle O\right\rangle $,
is given by 
\begin{eqnarray}
\delta\left\langle O\right\rangle  & = & \int d\mathbf{\Phi}\,\left[\frac{\overline{B}\left(\mathbf{\Phi}_{{\scriptscriptstyle \mathrm{B}}}\right)}{B\left(\mathbf{\Phi}_{{\scriptscriptstyle \mathrm{B}}}\right)}\,\Delta\left(\mathbf{\Phi}_{{\scriptscriptstyle \mathrm{B}}},\mathbf{\Phi}_{\mathrm{rad}}\right)-1\right]R\left(\mathbf{\Phi}_{{\scriptscriptstyle \mathrm{B}}},\mathbf{\Phi}_{\mathrm{rad}}\right)\,\left(O\left(\mathbf{\Phi}_{{\scriptscriptstyle \mathrm{B}}},\mathbf{\Phi}_{\mathrm{rad}}\right)-O\left(\mathbf{\Phi}_{{\scriptscriptstyle \mathrm{B}}}\right)\right)\label{eq:RESULTS-deltaO}
\end{eqnarray}
where the real phase space $\mathbf{\Phi}$, is factorised into that
of the underlying Born kinematics, $\mathbf{\Phi}_{{\scriptscriptstyle \mathrm{B}}}$,
and its complement parametrising those of the emitted parton, $\mathbf{\Phi}_{\mathrm{rad}}$.
$B$ and $\overline{B}$ refer to the leading order cross section
differential in the Born variables and, respectively, its NLO equivalent.
$R$ denotes the real cross section and $\Delta$ the Sudakov form
factor associated to emission from the underlying Born. From eq.~(\ref{eq:RESULTS-deltaO})
it is clear that \noun{Powheg }is NLO accurate for arbitrary infrared
safe observables, up to NNLO sized contributions; the factor containing
the difference of the observable functions in the rightmost bracket
allowing one to effectively replace the Sudakov form factor by 1.
In addition, it is also clear from eq.~(\ref{eq:RESULTS-deltaO}),
that the more inclusive the observable is with respect to the underlying
Born kinematics the greater will be the effect of the rightmost bracket
in nullifying $\delta\left\langle O\right\rangle $. Indeed, if the
observable is fully inclusive with respect to the Born kinematics
$\delta\left\langle O\right\rangle $ vanishes identically. The Higgs
$p_{{\scriptscriptstyle \mathrm{T}}}$ is not simply a function of
the Born kinematics, $ $$\mathbf{\Phi}_{{\scriptscriptstyle \mathrm{B}}}$,
as implemented in the \noun{Hj-Minlo }code, but it does have the property
that it is rather inclusive, allowing the rightmost bracket in eq.~(\ref{eq:RESULTS-deltaO})
to significantly diminish the difference between the \noun{Minlo }enhanced
fixed order prediction and that of the hardest emission cross section,
as seen in fig.~\ref{fig:pTH-HJ-MiNLO-NLO-vs-LH-vs-PY-vs-HNNLO}.

\bibliographystyle{JHEP}
\bibliography{nnlops}

\end{document}